# Re-inventing Willis

M.V. Simkin and V.P. Roychowdhury
*Department of Electrical Engineering, University of California, Los Angeles, CA 90095-1594*

Scientists often re-invent things that were long known. Here we review these activities as related to the mechanism of producing power law distributions, originally proposed in 1922 by Yule to explain experimental data on the sizes of biological genera, collected by Willis. We also review the history of re-invention of closely related branching processes, random graphs and coagulation models.

**Contents**

## 1. Introduction

The book of Ecclesiastes says: "Is there any thing whereof it may be said, See, this is new? It hath been already of old time, which was before us." As the job of the scientists is to discover new things, they are the most affected. In this paper we report a case study. The reader is likely familiar with the two concepts recently lauded as new: Preferential Attachment and Self Organized Criticality. Both were introduced as mechanisms of generating power-law distributions. It turns out that Preferential Attachment has been already of old time known as Yule's process, Simon's model or Cumulative Advantage. Self Organized Criticality may be indeed novel as a mechanism of tuning the system into a critical state. Apart from that, it is merely a branching process. Such process was discovered in mid 19th century and since then re-discovered at least six times. In fact, Preferential Attachment can also be considered a special kind of branching process and thus is closely related to Self Organized Criticality models. We also investigate the history of re-invention of closely related urn models, random graphs and coagulation processes and discuss the relation of Yule's process to the Renormalization Group.

## 2. Yule's process

In 1922 Willis and Yule [1] analyzed the data on frequency distribution of the sizes of biological genera, collected by Willis [2]. They discovered that this distribution follows a power law. To explain this observation Yule [3] proposed the following model. Consider two types of mutations: specific (that is producing a new specie of the same genus), which occur in each specie at rate $s$, and generic (that is producing a new genus) which occur in each genus at rate $g$.

In this model, the expectation value of the total number of genera grows with time as $\exp(g \times t)$. Therefore the probability distribution of the ages of genera is:

$$p(t) = g \exp(-gt). \qquad (2.1)$$

The expectation number of species in a genus of age $t$ is:

$$n(t) = \exp(s \times t). \qquad (2.2)$$

Now suppose that chance can be ignored, that the number of species in a genus can be taken as a continuous variable, and that the above can be taken as absolute functional relations. The size of genus is then absolutely determined by its age, and we can find the number of genera of each size by eliminating $t$ from Eq. (2.1) using Eq. (2.2):

$$t(n) = \frac{1}{s}\ln(n); \; dt = \frac{dn}{s \times n}.$$

This leads to:

$$p(n) = \frac{g}{s} n^{-\gamma}; \qquad \gamma = 1 + \frac{g}{s}. \qquad (2.3)$$

Yule [3] had also found the exact solution. The probability of a genus of an age $t$ to be monotypic, i.e. to consist of just one specie, is, obviously, $p_1(t) = \exp(-st)$. For a genus to contain two species at time $t$, a single mutation must occur at some intermediate time $t_1$, the original specie must not mutate until $t_1$, and two resulting species must not mutate for the time $t - t_1$. The probability for a genus to contain two species is obtained by integrating over all possible values of $t_1$:

$$p_2(t) = \int_0^t dt_1 p_1(t_1) s \exp(-2s(t-t_1)) =$$

$$\exp(-st)(1 - \exp(-st))$$

In general, one can verify by induction that:

$$p_n(t) = \exp(-st)(1 - \exp(-st))^{n-1} \qquad (2.4)$$

We see that the size distribution of genera of the same age is exponential. This result was rediscovered in 1992 by Günter *et al* [4] and in 2000 by Krapivsky and Redner [5], who used far more complicated mathematical methods.

Combining the distribution of the number of species in genera of age $t$ (given by Eq.(2.4)) with the distribution of the ages of genera (Eq.(2.1)) we obtain the overall probability distribution of genera with regard to the number of species in them:

$$p_n = \int_0^\infty dt\, p_n(t) p(t) =$$

$$\int_0^\infty dt \exp(-st)(1-\exp(-st))^{n-1} g \exp(-gt) =$$

$$\frac{g}{s} \int_0^1 dx\, x^{\frac{g}{s}} (1-x)^{n-1} = \qquad (2.5)$$

$$\frac{g}{s} \mathrm{B}\left(\frac{g}{s}+1, n\right) = \frac{g}{s} \frac{\Gamma\left(\frac{g}{s}+1\right)\Gamma(n)}{\Gamma\left(\frac{g}{s}+1+n\right)}$$

Here B and $\Gamma$ are Euler's Beta and Gamma functions. The large-$n$ asymptotic of Eq.(2.5) is

$$p_n \propto \frac{g}{s} \frac{\Gamma\left(\frac{g}{s}+1\right)}{n^{\frac{g}{s}+1}}. \qquad (2.6)$$

This derivation is simple. Surprisingly, one finds the following claim in modern literature [66] "Yule's analysis of the process was considerably more involved than the one presented here, essentially because the theory of stochastic processes as we now know it did not yet exist in his time." The given above solution, which follows that of Yule, is simpler than the one given in [66], which follows that given by Simon [11] (we will present it in Section 4).

To make the similarity between Yule's and Simon's model (which we will discuss in Section 4) more obvious we introduce the *modified* Yule's model. Let us assume that the rate of generic mutations is proportional to the number of species in a genus. In this case, the number of genera grows with time as $\exp((s+g)\times t)$ and the age distribution of genera is:

$$p(t) = (s+g)\exp(-(s+g)t). \qquad (2.7)$$

Substituting Eq.(2.7) into Eq.(2.5) we get:

$$p_n = \int_0^\infty dt\, p_n(t) p_t =$$

$$\int_0^\infty dt \exp(-st)(1-\exp(-st))^{n-1}(s+g)\exp(-(s+g)t)$$

$$= \frac{s+g}{s} \int_0^1 dx\, x^{\frac{s+g}{s}} (1-x)^{n-1} \qquad (2.8)$$

$$= \frac{s+g}{s} \mathrm{B}\left(\frac{g}{s}+2, n\right) =$$

$$\frac{s+g}{s} \frac{\Gamma\left(\frac{g}{s}+2\right)\Gamma(n)}{\Gamma\left(\frac{g}{s}+2+n\right)} \propto \frac{s+g}{s} \frac{\Gamma\left(\frac{g}{s}+2\right)}{n^{\frac{g}{s}+2}}$$

The major difference between the probability distribution, generated by the modified Yule's model (Eq.(2.8)) and the one generated by the standard Yule's model (Eq.(2.6)) is that Eq.(2.8) has the exponent of the power law equal to $\gamma = 2 + \frac{g}{s}$, while Eq.(2.6) has it equal to $\gamma = 1 + \frac{g}{s}$. By changing parameters $g$ and $s$, Eq.(2.6) can be tuned to have an exponent of any value greater than 1, while Eq.(2.8) has the exponent always greater than 2. Willis' data could be fitted with a power law with an exponent of around 1.5. Only standard Yule's model can account for this, while modified cannot.

## 3. Champernowne's model of income distribution

To explain the power law in distribution of incomes, discovered by Pareto, Champernowne [6] invented the following model. He divided income recipients into ranges of equal proportionate width. That is, if $I_{min}$ is the minimum income considered, then the first range contains persons with incomes between $I_{min}$ and $aI_{min}$, the second range includes persons with incomes between $aI_{min}$ and $a^2I_{min}$, and so on. Next he introduces transition probabilities $r_{nm}$ that a person who is in class $m$ at time $t$ will be in class $n$ at time $t + 1$. He assumes that $r_{nm}$ is a function of only $n$-$m$ (except for the small $m$, when it is modified to prohibit falling below minimum income).

To illustrate the Champernowne's model we will consider the simplest nontrivial transition function, the one that only allows transitions between adjacent classes:

$$r_{nm} = r(n-m) =$$
$$\begin{cases} r_1 & when \quad n-m=1 \\ r_0 & when \quad n-m=0 \\ r_{-1}=1-r_1-r_0 & when \quad n-m=-1 \end{cases}$$
$$\qquad\qquad\qquad\qquad (3.1)$$
$$r_{n0} = \begin{cases} r_1 & when \quad n=1 \\ r_0+r_{-1}=1-r_1 & when \quad n=0 \end{cases}$$

In equilibrium the occupation probabilities, $p_n$, of income ranges, $n$, should satisfy the following set of equations:

$$p_n = r_0 p_n + r_1 p_{n-1} + r_{-1} p_{n+1}$$
$$p_0 = (r_0 + r_{-1})p_0 + r_{-1}p_1 \qquad ,$$

which in the case $r_1 < r_{-1}$ has the following solution:

$$p_n = (1 - r_1/r_{-1}) \times (r_1/r_{-1})^n. \qquad (3.2)$$

The occupation probabilities decrease exponentially with income range number, $n$. As $p_n$ is the probability to have income between $a^{n-1}I_{min}$ and $a^n I_{min}$ - income exponentially grows with the range number. The situation is similar to what was in Yule's model, with time replaced by $n$. This leads to the income distribution of the form:

$$p(I) \sim I^{-\gamma}, \quad \gamma = 1 + \frac{\ln(r_{-1}/r_1)}{\ln(a)}. \qquad (3.3)$$

Champernowne's model was re-discovered in 1996 by Levy and Solomon [7].

## 4. Simon's model

The distribution of words by frequency of their use follows a power law. This fact was discovered sometime before 1916 by Estoup [8], re-discovered in 1928 by Condon [9], and once more in 1935 by Zipf [10]. Nowadays it is widely known as Zipf's law. To explain this observation Simon [11] proposed the following model. Consider a book that is being written, and that has reached a length of $N$ words. With probability $\alpha$ the $(N+1)$st word is a new word – a word that has not occurred in the first $N$ words. With probability 1-$\alpha$, the $(N+1)$st word is one of the old words. The probability that it will be a particular old word is proportional to the number of its previous occurrences.

If certain word appeared $K$ times among $N$ words, the probability that $(N+1)$st word will be that word is $(1-\alpha) \times \dfrac{K}{N}$. The evolution of the word frequency distribution (here $N_K$ denotes the number of words that appeared $K$ times) is described by the following equations:

$$\frac{dN_1}{dN} = \alpha - (1-\alpha) \times \frac{N_1}{N} \qquad (4.1)$$
$$\frac{dN_K}{dN} = (1-\alpha) \times \frac{(K-1) \times N_{K-1} - K \times N_K}{N}.$$

Assuming that the distribution of words has reached its stationary state, we can replace the derivatives with the ratios:

$$\frac{N_1}{N} = \alpha - (1-\alpha) \times \frac{N_1}{N} \qquad (4.2)$$

$$\frac{N_K}{N} = (1-\alpha) \times \frac{(K-1) \times N_{K-1} - K \times N_K}{N}.$$

The probability that the word occurs $K$ times is equal to

$$P_K = \frac{N_K(N)}{D(N)}, \qquad (4.3)$$

where $D(N) = \alpha N$ is the number of distinct words in the book.

We thus have:

$$N_K(N) = P_K \alpha N \qquad (4.4)$$

After substituting Eq. (4.4) in Eq. (4.2), we get:

$$P_1 = \frac{1}{2-\alpha},$$

$$\frac{P_K}{P_{K-1}} = \frac{K-1}{K + 1/(1-\alpha)} \quad (K > 1). \qquad (4.5)$$

Iterating the above equation we get

$$P_K = \frac{\Gamma(K) \Gamma(2 + 1/(1-\alpha))}{\Gamma(K+1+1/(1-\alpha))} \times \frac{1}{2-\alpha} =$$
$$\frac{\Gamma(K) \Gamma(1 + 1/(1-\alpha))}{\Gamma(K+1+1/(1-\alpha))} \times \frac{1}{1-\alpha} \qquad (4.6)$$

Eq. (4.6) has the following large $K$ asymptotic:

$$P_K \propto \frac{\Gamma(1 + 1/(1-\alpha))}{1-\alpha} \times K^{-\gamma}, \qquad (4.7)$$

where

$$\gamma = 1 + 1/(1-\alpha). \qquad (4.8)$$

Since the probability, $\alpha$, for the next word to be a new word is obviously small, obtained from Simon's model $\gamma$ is close to 2. This is about what one observes experimentally. Many publications dealing with the problem, including those by Condon and Zipf, use a different way of looking at the data - the rank-frequency representation. In this approach, one looks at the number of occurrences of a word, $K$, as a function of the rank, $r$, when the rank is determined by the above frequency of occurrence. One again finds a power law: $K(r) = C/r^{\delta}$. From rank-frequency distribution, one can find the number-frequency distribution, i.e. how many words appeared $K$ times. The number of words that appeared between $K_1$ and $K_2$ times is obviously $r_2-r_1$, where $K_1 = C/r_1^{\delta}$ and $K_2 = C/r_2^{\delta}$. Therefore, the number of words that appeared $K$ times, $N_K$, satisfies $N_K dK = -dr$ and hence, $N_K = -dr/dK \sim K^{-1/\delta-1}$. Therefore the exponent of number-frequency distribution, $\gamma$, is related to the exponent of rank-frequency distribution, $\delta$, as

$$\gamma = 1 + 1/\delta. \qquad (4.9)$$

This means that Simon's model gives for the rank-frequency distribution exponent: $\delta = 1-\alpha$. This means that $\delta$ is very close to 1, as is indeed seen in the experimental data.

In 1976, Price [12] used Simon's model to explain a power law distribution of citations to scientific papers, which he discovered in 1965 [13] (this was rediscovered in 1997 by Silagadze [14] and in 1998 by Redner [15]). He proposed to call it "cumulative advantage process". Simon's model was re-discovered in 1992 by Günter *et al* [4] and 1999 by Barabasi and Albert [16]. In the latter case, it acquired the new name: "preferential attachment".

## 5. Solution of Simon's model by Yule's method

If $M$ is the number of appearances of a particular word, and N is the total number of words, than the probability that the number of occurrences of this word will increase by one when the next word is added to the sequence is:

$$p(M \rightarrow M+1) = (1-\alpha)\frac{M}{N}.$$

This can be rewritten as:

$$p(M \rightarrow M+1) = (1-\alpha)\frac{M}{N}dN,$$

where $dN = 1$. By introducing the variable $t = \ln(N)$, we get:

$$p(M \rightarrow M+1) = (1-\alpha)Mdt.$$

Note, that because $dt = \frac{dN}{N}$, $t$ is a continuous variable in the limit of large $N$. Similarly, the probability that the number of distinct words increases by one when the next word is added to the sequence is:

$$p(D \rightarrow D+1) = \alpha dN = \alpha N \frac{dN}{N} = \alpha N dt.$$

If a word corresponds to specie and a distinct word to a genus, than Simon's model is equivalent to the modified Yule's model with the rate of specific mutations equal $1-\alpha$, and the rate of generic mutations equal $\alpha$.

## 6. Solution of Yule's model by Simon's method

We will start with the *modified* Yule's model. In Simon's model genus corresponds to a distinct word, and the number of species in a genus corresponds to the number of occurrences of a word in the text. The probability for the new mutation to be generic corresponds to the probability for the next word to be a new word. It is equal to:

$$\alpha = \frac{g}{s+g}.$$

Substituting this into Eq. (4.6), we recover Eq. (2.8).

Original Yule's model is more difficult to solve by Simon's method. The problem is that the probability of a new mutation to be generic changes in time. This probability is given by the equation:

$$\alpha = \frac{gN_g}{sN_s + gN_g}, \tag{6.1}$$

where $N_g$ is the total number of genera, and $N_s$ the total number of species. Let us compute these numbers at time $t$. Suppose, that at time 0 there was one genus, consisting of a single specie. The expectation number of genera at time $t$ is, obviously,

$$N_g = \exp(g \times t). \tag{6.2}$$

The expectation number of species in the primal genus at time $t$ is:

$$N_s^1 = \exp(s \times t)$$

The number of species in new genera is:

$$N_s^* = \int_0^t dt_1 g \exp(gt_1) \exp(s(t-t_1)) =$$

$$\frac{g}{g-s}(\exp(gt) - \exp(st))$$

The expectation number of all species at time $t$ is:

$$N_s = N_s^1 + N_s^* = \frac{g}{g-s}\exp(gt) + \frac{s}{s-g}\exp(st)$$

The large $t$ asymptotic of the above is:

$$N_s = \begin{cases} \dfrac{g}{g-s}\exp(gt) & when \quad g > s \\ \dfrac{s}{s-g}\exp(st) & when \quad s > g \end{cases} \qquad (6.3)$$

Let us consider the case $g > s$. Substituting Eqs.(6.2) and (6.3) into Eq.(6.1) we get:

$$\alpha = \frac{ge^{gt}}{s\dfrac{g}{g-s}e^{gt} + ge^{gt}} = 1 - \frac{s}{g} \qquad (6.4)$$

Substituting Eq.(6.4) into Eq.(4.4) we recover Eq.(2.5).

Let us now consider the case $s > g$. After substituting Eqs.(6.2) and (6.3) into Eq.(6.1) we get:

$$\alpha(t) = \frac{ge^{gt}}{s\dfrac{s}{s-g}e^{st} + ge^{gt}} \propto g\frac{s-g}{s^2}e^{-(s-g)t} \qquad (6.5)$$

After expressing $t$ through $N_s$, using Eq.(6.3), and substituting the result into Eq.(6.5) we get:

$$\alpha(N_s) = CN_s^{-\gamma}; \quad \gamma = 1 - \frac{g}{s}. \qquad (6.6)$$

Here $C$ is a function of $g$ and $s$.

Let us consider a modified Simon's model where

$$\alpha = CN^{-\gamma}. \qquad (6.7)$$

The number of distinct words as a function of total number of words will be

$$D(N) = \int_0^N dMCM^{-\gamma} = \frac{C}{1-\gamma}N^{1-\gamma} \qquad (6.8)$$

After substituting Eq.(6.8) into Eq.(4.3) and the result together with Eq.(6.7) into Eq.(4.2) we get:

$$P_1 = \frac{1-\gamma}{2-\gamma}$$

$$\frac{P_K}{P_{K-1}} = \frac{K-1}{K+1-\gamma} \qquad (6.9)$$

By iteration of Eq. (6.9), we get:

$$P_K = (1-\gamma)\frac{\Gamma(K)\Gamma(2-\gamma)}{\Gamma(K+2-\gamma)} \qquad (6.10)$$

after substituting $\gamma = 1 - \dfrac{g}{s}$ into Eq.(6.10) we recover Eq.(2.5). This model with $N$-dependent $\alpha$ was first suggested and solved by Simon [11]. It was rediscovered by Dorogovtsev and Mendes [17] in context of the science of networks (Section 15).

This exercise shows that Yule's and Simon's models are two ways of looking at the same thing. In contrast, Champernowne's model is similar, but not identical.

# 7. Markov-Eggenberger-Polya Urn models

In 1907 Markov was wondering what happens with the law of large numbers when the variables are dependent [18]. Consider an urn with one white and one black ball. Let us pull a random ball out of the urn, record its color and put the ball back. If we make a large number $N$ of such independent trials, the law of large numbers tells us that the fraction of pulled out white balls is on average 0.5 and the standard deviation from this average is of the order $1/\sqrt{N}$. Thus expectation value becomes exact value in the limit of large $N$. Markov modified the model in the following way. The urn initially contains one white and one black ball. We pull out a random ball, then put it back and in addition add to the urn another ball of the same color. We repeat the procedure again and again. Trials are no longer independent but the outcome of the trial depends on the outcomes of all preceding trials. Obviously, the procedure has cumulative advantage feature of Yule-Simon process. This connection was pointed out in 1975 by Price [12], and re-discovered in 2003 by Chung, Handjani, and Jungreis [20]. Markov's problem is easier to solve than Yule's. After one step there can be in the urn either one black and two white balls or one white and two black balls with equal probability. It is easy to show that after two steps the possible combinations are 3-1, 2-2, and 1-3, and all of them have the same probability. After $N$ steps the urn can contain any number of white balls from 1 to $N+1$ and all these realizations have equal probabilities $1/N$. This can be proved by induction. If this holds true after $N$ steps, then the probability to have $n$ white balls after $N+1$ steps is

$$p(n, N+1) = \frac{n-1}{N+1} p(n-1, N) + \frac{N+1-n}{N+1} p(n, N) =$$

$$\frac{N}{N+1} \frac{1}{N} = \frac{1}{N+1}$$

The distribution of white balls in the urn after $N$ steps is uniform between 1 and $N+1$. This means that the distribution of the number of pulled out white balls is uniform between 0 and $N$. Thus, the expected value of the fraction of pulled out white balls is 0.5 just as it was in the simplistic model. However, the standard deviation is of order unity. Thus, the standard law of large numbers no longer holds when the events are dependent.

Markov's model was re-invented in 1923 by Eggenberger and Polya (see Ref. [19] pp.176-177) and is widely known today as Polya's urn scheme.

The resulting uniform distribution of balls in the urn is almost identical to the distribution of bosons between the two quantum states, corresponding to a doubly degenerate energy level. The only difference is as follows. The probability of adding a ball of a given color is proportional to the current number of balls of this color in the urn. For bosons, the probability of transition into a particular state is proportional not to the present number of bosons in this state, but to this number plus one. Therefore, the distribution of *added* balls (excluding the original two balls in the urn) is exactly given by Bose statistics. We do not know if there exists any interesting physical system with these properties, but a modified Markov's model, where the urn initially contains three balls of three different colors, has physical realization showing an interesting new effect. This is a Bose-Einstein condensate of spin-1 bosons. We will investigate the system in more detail in Section 16. Here will just note that similar to the two level case, where the expected excess number of white balls over black balls is not of the order $\sqrt{N}$ but is of the order $N$, the excess number of particles with $s_z = 1$ over the number of particles with $s_z = -1$ is of the order $N$. This means that Bose-Einstein condensate shows a spontaneous magnetization, or, in other words, is ferromagnetic. See Refs [86] and [87].

There is another way to modify Markov's urn scheme. With probability $1-\alpha$ we just do as before but with probability $\alpha$ we add a ball of a new color. Now we get exactly Simon's process. Such modification of Markov's model was proposed in 2003 by Chung, Handjani, and Jungreis [20]. Their model, however, is identical to the one proposed much earlier by Ijiri and Simon [22] in connection with Bose-Einstein statistics (see Section16).

## 8. Genetic model of Moran

In 1958 Moran [43] introduced the following model. There is a gene pool of fixed but large size. At each step one selected at random gene dies. To replace it we select a random gene from the pool and add a gene of the same type. With the probability $\alpha$ the added gene can mutate.

The model looks similar to Simon's model and can be solved using the same method. Equilibrium gene frequencies should satisfy the following equations (here $N_K$ denotes the number of genes that appeared $K$ times):

$$-\frac{N_1}{N} + \frac{2N_2}{N} - (1-\alpha) \times \frac{N_1}{N} + \alpha = 0 \qquad (8.1)$$

$$-\frac{K \times N_K}{N} + \frac{(K+1) \times N_{K+1}}{N} -$$

$$(1-\alpha) \times \frac{K \times N_K}{N} + \qquad (K>1) \qquad (8.2)$$

$$(1-\alpha) \times \frac{(K-1) \times N_{K-1}}{N} = 0$$

To find $N_1$ note that the rate of creation of new genes is $\alpha$. At the same time the rate of loss of distinct genes is $\frac{N_1}{N}$ : clearly, a gene gets extinct when a gene which occurs only once is selected to die. It follows that $\frac{N_1}{N} = \alpha$. Substituting this in Eq.(8.1), we get $\frac{N_2}{N} = \frac{\alpha(1-\alpha)}{2}$ , and, similarly, $\frac{N_3}{N} = \frac{\alpha(1-\alpha)^2}{3}$ . In general, it can be verified by induction that

$$\frac{N_K}{N} = \frac{\alpha(1-\alpha)^{K-1}}{K} . \qquad (8.3)$$

The probability that the gene occurs $K$ times is equal to

$$P_K = N_K / D , \qquad (8.4)$$

where $D$ is the number of distinct genes in the pool. The latter can be computed as

$$D = \sum_{K=1}^{\infty} N_K = N \frac{\alpha \ln(1/\alpha)}{1-\alpha} \qquad (8.5)$$

After substituting Eq.(8.5), and into Eq.(8.4) we get:

$$P_K = \frac{(1-\alpha)^K}{\ln(1/\alpha)K} . \qquad (8.6)$$

The gene frequency distribution follows a hyperbolic law with an exponential cut-off.

Moran's model was first solved for gene frequency distribution by Karlin and McGregor [44] who used a far more complicated method. They solved the model exactly for a finite $N$. Equation (8.3) can be obtained by taking the limit $N \to \infty$ in Eq.(3.7) of Ref.[44].

Moran's model can be reformulated in verbal terms. Consider a string of words of large but fixed length $N$. At each step we delete one randomly selected word and add one word according to the rules of Simon's model. (With probability $\alpha$ it is a new word. With probability 1-$\alpha$ it is one of the old words. The probability that it is a particular old word is proportional to the number of its previous occurrences.)

A small modification of Moran's model makes it very similar to Simon's model. The rules of word addition remain the same. The rule of deletion will be as follows. The deletion happens with probability $\alpha$. We select a random *distinct* word (the probability is equal for all distinct words in the sequence and does not depend on the number of occurrence of the word). Then we delete all occurrences of this word from the sequence. With these new rules the equilibrium equations (8.1), (8.2) will change only slightly. Third and fourth terms will not change, as the rules of addition didn't change. Second term will disappear, because if a word is deleted then all its occurrences are deleted, it does not just move from $N_K$ to $N_{K-1}$ category, as it was in Moran's model. The first term will become $-\alpha \frac{N_K}{D}$, where $D$ is the number of distinct words in the sequence. To compute $D$ note that the average number of occurrences of a word is $N/D$. In equilibrium, the rate of addition must be equal to the rate of deletion. Therefore $\alpha N/D = 1$, or $D = \alpha N$. The first term becomes $\frac{N_K}{N}$, and Eqs.(8.1,2) transform into:

$$-\frac{N_1}{N} - (1-\alpha) \times \frac{N_1}{N} + \alpha = 0 ,$$

$$-\frac{N_K}{N} + (1-\alpha) \times \frac{(K-1) \times N_{K-1}}{N} -$$

$$(1-\alpha) \times \frac{K \times N_K}{N} = 0 \qquad (K > 1) \qquad (8.7)$$

The above equations are identical to Eq. (4.2) and therefore have the same solution, Eq. (4.6). Although the above model has the same solution as standard Simon's model, it is not identical to it. In the latter words are added but not deleted and the number of words in the sequence is growing. In the above model words are both added and deleted and the number of words in the sequence is constant. This model was formulated and solved by Simon in the same paper were he introduced his better-known model [11].

## 9. Spectrum of cosmic radiation

In 1949 Fermi [24] explained experimentally observed power-law spectrum of cosmic radiation as follows. The particles, like protons, are accelerated through collisions with wandering interstellar magnetic fields. An elementary estimate can be obtained by picturing the collisions against reflecting obstacles of very large mass, moving with random velocities averaging to $V$. Assuming this picture, one finds that the average gain in energy per collision is given as order of magnitude by

$$dE = \left(\frac{V}{c}\right)^2 E,$$

where $E$ represents the energy of the particle inclusive of rest energy, and $c$ is the speed of light. If we call $\tau$ the time between scattering collisions, the energy acquired by a particle of age $t$ will be

$$E(t) = Mc^2 \exp\left(\left(\frac{V}{c}\right)^2 \frac{t}{\tau}\right).$$

During the process of acceleration, a proton may lose most of its energy by a nuclear collision. This absorption process can be considered to proceed according to an exponential law. We expect the age probability distribution to be

$$p(t) = \exp\left(-\frac{t}{T}\right),$$

where $T$ is the time between absorption collisions. The problem is identical to Yule's with particle equivalent to genera and energy to the number of species. Combining relationships between age and energy with the probability distribution of age, we find the probability distribution of the energy:

$$p(E) \propto E^{-1-\frac{\tau}{T}\left(\frac{c}{V}\right)^2}$$

## 10. Renormalization group

Near critical temperature, $T_c$, of a second order phase transitions physical parameters of the system are power-law functions of the reduced temperature, $t = \frac{T - T_c}{T_c}$. In 1971 Wilson [25] developed the renormalization group (RG) method to explain this phenomenon. He studied how parameters of the system change after $n$ successive re-scalings of the length by a factor $l$. He found that the reduced temperature grows exponentially with $n$: $t^{(n)} = \left(l^{y_t}\right)^n t$. At the same time correlation length decreases exponentially with $n$: $\xi^{(n)} = l^{-n}\xi$. As a result the correlation length scales with the reduced temperature as: $\xi(t) \propto t^{-1/y_t}$. Similar to what happened in Yule's model, the power law came out of two exponential dependencies.

So far, we derived all of the equations used in the article. However, Wilson's work is too complicated to discuss here. Therefore, we will illustrate RG on a simple example of the so-called percolation problem. The problem was originally formulated for liquids percolating through a porous media, a question of interest in oil production. Afterward the research in the field shifted into studying lattice models, which are abstract and remote from reality, but, nonetheless, very good for keeping scientists busy. They consider a lattice with some nodes removed at random and fraction *p* of the nodes left. Just like the triangular lattice shown in Figure 1 (a), where we show removed nodes as empty circles and remaining nodes as black ones. The question of interest is what is the critical value $p_c$ of the fraction of remaining nodes at which an infinite cluster of connected nodes emerges. One way to answer it is to try to look at the lattice from a distance. When we are sufficiently far away, we cannot distinguish separate nodes but only blocks of three nodes. These blocks are shown as grey triangles. If the block has three or two black nodes then the whole block will appear to us black. If it has one or zero black nodes it will appear to us white. Figure 1 (b) shows what becomes with lattice shown in Figure 1 (a) after the just described block transformation. The renormalized value $p'$ of the fraction of remaining nodes is given by the following equation

$$p' = p^3 + 3p^2(1-p)$$

Here the first term is the probability that all of the three nodes in the block are black and the second term is the probability that there are exactly two black nodes in the block. The above equation has three fixed points, for which $p' = p$. They are 0, ½, and 1. It is easy to see that if *p* is close to 0, then $p'$ will be even closer to 0. Similarly, if *p* is close to 1, then $p'$ will be even closer to 1. In other words, these fixed points are stable. In contrast, the fixed point $p = 1/2$ is unstable. When *p* is away from ½, then $p'$ is even further. We can in turn apply the block transformation to the renormalized lattice and do it again and again. The flow diagram in Figure 2 illustrates the evolution of *p* under the renormalization group transformation. When we start with $p > 1/2$, after repeated RG transformations we will get to the fixed point $p = 1$, which corresponds to a fully occupied lattice and indicates that there is an infinite connected cluster.

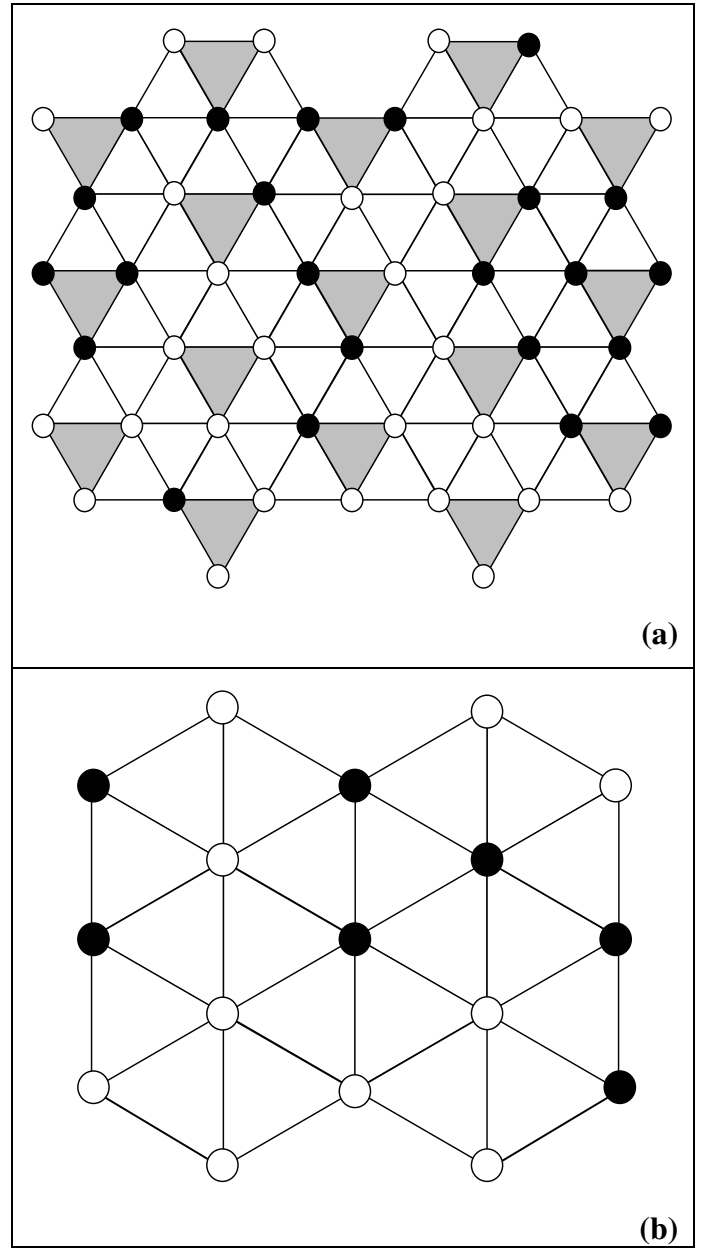

**Figure 1.** Renormalization Group transformation for a triangular lattice.

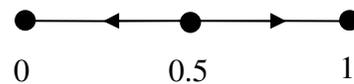

**Figure 2** Renormalization group flows for percolation on the triangular lattice.

If we start with $p < 1/2$, after repeated RG transformations we will get to the fixed point $p = 0$, which corresponds to an empty lattice. Thus $p_c = 1/2$ is the critical point, which separates these two regimes[1].

Apart from the value of $p_c$ an important parameter is the size $\xi$ of the largest connected cluster for $p < p_c$. When $p$ is close to $p_c = 1/2$, the renormalization equation can be expanded as

$$p' - p_c = 3/2\left(p - p_c\right).$$

After $n$ such transformations we get

$$p^{(n)} - p_c = \left(3/2\right)^n \left(p - p_c\right)$$

The length is reduced by the factor of $\sqrt{3}$ with each RG transformation. Thus $\xi^{(n)} = \left(1/\sqrt{3}\right)^n \xi$. From this follows:

$$\xi(p) \sim \left(p - p_c\right)^{-\nu},$$

where $\nu = \ln\left(\sqrt{3}\right)/\ln(3/2) \approx 1.35$.

This pedagogical example of RG was first published by Reynolds, Klein and Stanley [45] and rediscovered by Thouless [46] and once more by Stinchcombe [47]. However, in his later work [48] Stinchcombe cited both of his predecessors. This is were we learned of them from.

## 11. Bradford's law

Bradford Law states that the distribution of scientific journals by the number of articles, they contain, follows a power law. In 1970 Naranan [26] considered the model where both the number of journals and the size of each journal increase exponentially with time. He obtained an equation identical to Eq. (2.3) with $g$ and $s$ being growth rates of the number of journals and number of articles in a journal respectively. This very mechanism was rediscovered in 1999 by Huberman and Adamic [28] in the context of distribution of websites by the number of webpages they contain. Since all journals now have websites and an article is a webpage on journal's website, one can say that Huberman and Adamic studied almost the same system as Naranan. Both aforementioned reinventions ([26], [28]) appeared in the same journal, *Nature*, which published the original paper by Willis and Yule [1].

In his first paper on this topic [26] Naranan mentioned the 1949 paper by Fermi [24] as the source of the idea. Later he discovered [27] that the original idea was much older and that it was several times re-invented. He gave a list of related papers, which helped us in writing this article.

## 12. Psychophysical law

Perceived intensity of physical stimulus varies with their physical intensity as a power law [33]. For example, for brightness the exponent is about 1/3. That is if we increase the luminous power of the light source 8 times it will appear only two times brighter. This behavior is puzzling since frequency of nerve impulses from sensory receptors is proportional to a logarithm of the intensity of stimulations. In 1963, MacKay [34] proposed the matched response model, which can resolve the apparent paradox. In his model, perception is an adaptive response to stimulation. Perceived intensity of a stimulus reflects not the frequency of impulses from receptor organ, but the magnitude of internal activity evoked to match or counterbalance that frequency. MacKay proposed that frequency of matching impulses is logarithmically related to the magnitude of internal activity. In such case both perceived intensity and physical intensity depend exponentially on the frequency of nerve impulses and we get a power law relation between them, similarly to how we got it in Yule's model.

---

[1] Interestingly this result is exact as can be proven by duality argument [46]. Every finite cluster of black nodes is surrounded by a perimeter of nearest-neighbor white nodes, which are themselves connected. Thus for every finite cluster of black nodes there is a larger cluster of white nodes. At the percolation threshold there is no limit on the size of black clusters and from the above argument it follows that there is also no limit on the size of white clusters. However, percolation for white nodes is dual to percolation for black nodes. Thus the critical point for white nodes percolation is $1 - p_c$. We thus have $1 - p_c = p_c$ or $p_c = 1/2$.

# 13. Optimization of monkey-typed text

In 1953 Mandelbrot [31] explained Zipf's law in word frequencies by showing that such distribution maximizes Shannon's information entropy. We transmit information using words, which are coded using sequences of symbols. Suppose we have $q$ symbols which occur with probabilities $p_g$ and have costs $c_g$. We want to maximize the ratio of information entropy

$$H = -\sum p_g \ln(p_g) \qquad (13.1)$$

to cost

$$C = \sum p_g c_g \qquad (13.2)$$

under the obvious restriction

$$\sum p_g = 1. \qquad (13.3)$$

So we need to maximize $\dfrac{H}{C} - \lambda \sum p_g$, where $\lambda$ is a Lagrange multiplier. The result is

$$p_g = e^{-(1+\lambda C) - \frac{H}{C} c_g}$$

To satisfy Eq.(13.2) we need to set $1 + \lambda C$ to 0. We thus have:

$$p_g = e^{-\frac{H}{C} c_g} \qquad (13.4)$$

Next we need to substitute Eq.(13.4) into Eq.(13.3) and solve it for $H/C$. This is easy to do when all costs are equal $c_g = c$. Then we get

$$q e^{-\frac{H}{C} c} = 1 \qquad (13.5)$$

or $\dfrac{H}{C} = \dfrac{\ln(q)}{c}$. Consequently $p_g = \dfrac{1}{q}$. That is information entropy is maximum when all symbols are equally probable.

However, we need a space between words and for that a special symbol. Suppose its cost is $c_{sp}$. Then instead of Eq.(13.5) we get

$$q e^{-\frac{H}{C} c} + e^{-\frac{H}{C} c_{sp}} = 1. \qquad (13.6)$$

This is in general a transcendental equation. It is easy to solve in the case $c_{sp} = c$, what Mandelbrot did. We will slightly deviate from his exposition and will tackle the general case to make the connection with the subsequent work of Miller [29] more transparent. Instead of solving the transcendental Eq.(13.6) we will express our results in terms of the probability of space, $p_{sp}$[2]. Since the costs of all letters are identical, their probabilities, according to Eq.(13.4), are also identical and, according to Eq.(13.3), should be equal to $p = \dfrac{1 - p_{sp}}{q}$.

Now we will finally compute the word frequencies distribution. The probability of any given $n$-letter word is

$$P(n) = p^n p_{sp} = \left(\frac{1 - p_{sp}}{q}\right)^n p_{sp} = p_{sp} e^{-n \ln\left(\frac{q}{1 - p_{sp}}\right)} \qquad (13.7)$$

The number of $n$-letter words is

$$N(n) = q^n = e^{n \ln(q)} \qquad (13.8)$$

From this, we need to find the number of words occurring with given probability $N(P)$. The task identical to that encountered in Yule's problem (Section 2) with $n$ analogous to time and, therefore, $-n$ to age. We get $N(P) = P^{-\gamma}$ with

$$\gamma = 1 + \frac{\ln(q)}{\ln(q) - \ln(1 - p_{sp})} \qquad (13.9)$$

---

[2] One can express $c_{sp}$ through $p_{sp}$ as $c_{sp} = c \ln(p_{sp}) / \big(\ln(1 - p_{sp}) - \ln(q)\big)$ and use this equation to compute $p_{sp}$ for given $c_{sp}$ numerically. We are not going to do this, anyway, since the observable parameter is $p_{sp}$.

Using Eq (4.9) we can get from Eq.(13.9) the exponent of the rank-frequency distribution

$$\delta = 1 - \frac{\ln(1 - p_{sp})}{\ln(q)} \ . \qquad (13.10)$$

In 1957 Miller [29] had proposed a radically different explanation of Zipf's law in word frequencies. A monkey, sitting at a typewriter, is randomly hitting its keys. The resulting discourse is a random sequence of independent symbols: the letters and the space that mark the boundaries between words. There are $q$ equiprobable letters and space, which has the probability $p_{sp}$. The reader who had read the present section from the beginning immediately realizes that Miller's explanation is identical to Mandelbrot's. Not surprisingly, he obtained Eqs (13.7) and (13.8) and from them derived Eq.(13.10). Substituting into Eq.(13.10) the values $q = 26$ and $p_{sp} = 0.18$ [3], which he took from English language, Miller got $\delta = 1.06$, in good agreement with the experimental data.

In his paper [29] Miller insisted that his explanation is different from that of Mandelbrot, since he did not optimize the ratio of information to cost. As we had seen, the only optimization used by Mandelbrot was to show that the probabilities of symbols are identical when their costs are identical. The use of Eq. (13.1) implies that the symbols are random. It only gives the correct value of information entropy

---

when symbols are uncorrelated [84] (otherwise it is only an upper bound). Completely random sequence of equiprobable symbols maximizes Shannon's information. Therefore, Miller did indeed maximize information, without knowing it. His model is, thus, identical to that of Mandelbrot. In view of that, it may seem unreasonable to devote so much space to Mandelbrot's calculations. We did it because Mandelbrot's paper is sometimes referred to as "a classical and beautiful model by Mandelbrot" [85]. Some authors repeat Miller's mistake that his model is different from Mandelbrot's even today. For example, Mitzenmacher [67] writes, "A potentially serious objection to Mandelbrot's argument was developed by the psychologist Miller who demonstrated that the power law behavior of word frequency arises even without an underlying optimization problem." In contrast, Mandelbrot understood that the two explanations of Zipf's Law are identical (see Ref. [32]).

Miller's explanation was rediscovered in 1992 by Li [30]. By the way, Li credited Miller for the idea, citing his Introduction to the book by Zipf [10], where Miller wrote that typing monkeys will produce Zipf's law, but did not give a mathematical proof. Apparently, Li decided that Miller never produced that proof.

## 14. HOTly designed systems

In 1999 Carlson and Doyle [101] introduced the concept of Highly Optimized Tolerance (HOT) according to which power laws can arise because of optimal design to tolerate failures. The toy model they considered was that of forest fires. Suppose that there is a forest, different regions of which have different probability to be struck by a lightning and catch fire. The forest service wants to minimize the damage from fire by constructing firebreak roads. However, it costs some money per mile to maintain fire roads. Therefore, there is a trade off between the loss from fire and the cost of fire roads maintenance. In the region of higher density of firebreaks, the fires will have smaller sizes. Therefore, the fire road system should have higher density in the regions where a fire can start with higher probability. Now we proceed to derive the distribution of the sizes of forest fires. Suppose that in certain region fire roads divide the forest into squares with side $l$. The cost of losses from fire $C$ is proportional to burnt area $A = l^2$, that is $C = c_A A$, where $c_A$ is some proportionality constant. The cost of road maintenance $R$ is proportional to their combined length, that is $R = c_R/l = c_R A^{-1/2}$. To find the optimum value of $l$ (or $A$) we should minimize the total expected cost

$$E = pC + R = pc_A A + c_R A^{-1/2} \qquad (14.1)$$

where $p$ is the probability that a fire starts in the area. Expected cost given by Eq.(14.1) reaches its minimum when $A = (c_R/2c_A)^{2/3} p^{-2/3}$. If we know the probability distribution of $p$ we can compute the probability distribution of $A$.

Meanwhile Carlson and Doyle propose a generalization of the model. Consider some general system susceptible to failures. The cost of failure of size $A$ is given by

$$C = c_A A^\alpha \qquad (14.2)$$

The cost of recourses necessary to restrict the failure to size $A$ is

$$R = c_R A^{-1/\beta} \qquad (14.3)$$

By minimizing the analog of Eq.(14.1) we get

$$A \sim p^{-1/\gamma}, \qquad (14.4)$$

where $\gamma = \alpha + 1/\beta$.

Now suppose that there are many sites (which we will index using a variable $n$) where the system can have a failure and each site has the probability of failure $p(n)$. The size of a failure at site $n$ is

$$A(n) \sim [p(n)]^{-1/\gamma} \qquad (14.5)$$

Carlson and Doyle considered several different forms of $p(n)$. One of them was

$$p(n) \sim e^{-n} \qquad (14.6)$$

After substituting Eq.(14.6) into Eq.(14.5) we get

$$A(n) \sim e^{n/\gamma}$$

Now we get a power law distribution of failure sizes out of two exponents, just as it was in Yule's model:

$$p(A) \sim A^{-1-\gamma}.$$

In Table 14.1 we give the results for two more versions of $p(n)$ considered by Carlson and Doyle. The are obtained using the equation

$$p(A) = p(n)/(dA/dn) . \qquad (14.7)$$

Up to this point, we just followed the exposition of Carlson and Doyle apart from simplifying some mathematical derivations. Now it is time to ask questions. Actually, we made one more change. What we called an index and denote as $n$, Carlson and Doyle called a coordinate and denote as $x$. Why should failure probability be a monotonously decreasing function of a coordinate? This does not make any sense in the case of a forest or in any other conceivable case. The only way to make sense of it is to interpret $n$ as a rank and $p(n)$ as a Zipfian distribution. So that the most failure prone site has $n = 1$, the next $n = 2$ and so on. We already derived the number distribution from rank distribution in a particular case when rank distribution was a power law: see the paragraph preceding Eq.(4.9). It is easy to do in the general case. We get:

$$N(p) = -1/(dn/dp), \qquad (14.8)$$

where $N(p)$ is the number of sites which have failure probability $p$. After substituting Eq.(14.6) into Eq.(14.8) we get

$$N(p) \sim 1/p$$

Results for two more cases are shown in Table14.1. Now we see that to get a power law of failure sizes Carlson and Doyle had to postulate three other power laws: Eqs.(14.2), (14.3) and any one of those in right column of Table 14.1.

**Table 14.1** Failure probability, $p$, as a function of site rank $n$ - $p(n)$. Number of sites, $N$, with failure probability $p$ - $N(p)$. Probability distribution of failure sizes - $p(A)$.

| $p(n)$ | $p(A)$ | $N(p)$ |
|--------|--------|--------|
| $n^{-q}$ | $A^{-1-\gamma+\gamma/q}$ | $p^{-1-1/q}$ |
| $e^{-n}$ | $A^{-1-\gamma}$ | $p^{-1}$ |
| $e^{-n^2}$ | $A^{-1-\gamma}[\ln(A)]^{-1/2}$ | $p^{-1}[\ln(p_{max}/p)]^{-1/2}$ |

Let us now derive a formula to get $p(A)$ from $N(p)$. The probability $\tilde{p}(p)$ of a failure at one of the sites with failure probability $p$ is equal to

$$\tilde{p}(p) = pN(p).$$ (14.9)

Distribution of failure sizes can be computed as

$$p(A) = \tilde{p}(p)/(dA/dp)$$ (14.10)

After substituting Eqs (14.4) and (14.9) into Eq.(14.10) we get

$$p(A) \sim A^{-1-2\gamma}N\left(A^{-\gamma}\right)$$ (14.11)

This means that any distribution $N(p)$ which does not vanish at $p = 0$ will produce the large $A$ asymptotic $p(A) \sim A^{-1-2\gamma}$. So we do not have to postulate a third power law after all, just two are sufficient.

## 15. The Science of Networks

In 1999 in order to explain the power-law distribution of the connectivity in the World Wide Web [35] and other networks Barabasi and Albert [16] proposed the Preferential Attachment model. Starting with a small number of nodes, at every time step we add a new node and link it to one of the nodes already present in the system. When choosing the nodes to which the new node connects, we assume that the probability $p_i$ that a new node will be connected to node $i$ is proportional to the number of nodes already linking to it (its degree) $k_i$. After $t$ time steps there are $t$ nodes in the network. The average degree is equal to 2 (we add one link with every node, but each link connects 2 nodes). Therefore: $p_i = \dfrac{k_i}{2t}$. If we assume that $t$ can be treated as a continuous variable, the expectation number for the degree of any given node obeys the following evolution equation: $\dfrac{d\langle k_i \rangle}{dt} = p_i(t) = \dfrac{\langle k_i \rangle}{2t}$. The solution of this equation, with the initial condition that the node, $i$, at the time of its introduction, $t_i$, has $k_i(t_i) = 1$, is $\langle k_i \rangle = \left(\dfrac{t}{t_i}\right)^{1/2}$.

Assuming that chance can be ignored and that the above equation is the exact functional relation between $k_i$ and $t_i$ we can easily compute the degree distribution. The nodes introduction times are uniformly distributed between 0 and $t$, therefore:

$$p(t_i) = \frac{1}{t}.$$ For the degree distribution, we have:

$$p(k_i) = p(t_i)\frac{dt_i}{dk_i} = \frac{1}{t}2\frac{(t_i)^{3/2}}{t^{1/2}} = \frac{2}{(k_i)^3}.$$

In this derivation, we assumed that the network grows at constant speed. In reality many networks (like WWW) grow exponentially with time. Thus $t$ (which is also the total number of nodes) can be expressed through real time, $\tau$, as $t = e^\tau$. Substituting this into the evolution equation we get $\dfrac{d\langle k_i \rangle}{de^\tau} = \dfrac{\langle k_i \rangle}{2e^\tau}$ or $\dfrac{d\langle k_i \rangle}{d\tau} = \dfrac{\langle k_i \rangle}{2}$, which has the solution $\langle k_i \rangle = e^{\tau/2}$. We recognize here the modified Yule's model with node being a genus, degree – number of species in a genus, and the mutation rates: $g = s = \frac{1}{2}$. The Barabasi-Albert model is thus identical to Yule's model and their solution differs from Yule's solution by mere variable substitution.

Although solution used by Barabasi and Albert is very similar to that used by Yule, the formulation of their model of Preferential Attachment is closer to that of Simon. With the substitution *node = distinct word* and *degree = number of occurrences* Barabasi-Albert model reduces to Simon's model with $\alpha=1/2$. This was pointed out in [36].

## 16. Cumulative advantage and Bose-Einstein statistics

In 1974 Hill [21] pointed out the connection between Bose statistics and Yule-Simon process. In 1977 Ijiri and Simon further discussed it (see the Chapter "Some distributions associated with Bose-Einstein statistics" in Ref. [22]). The link was rediscovered in 2001 by Bianconi and Barabasi [23].

Let us consider a system of $N$ bosons sitting on an $L$-fold degenerate energy level. Our aim is to find the probability distribution of the quantum states' occupancies. From probability theory perspective, this is a problem of distributing $N$ indistinguishable balls in $L$ distinguishable bins. Here balls correspond to $N$ bosons and bins to $L$ quantum states corresponding to the $L$-fold degenerate energy level. Let us first compute the number of distinguishable arrangements, $D(N, L)$. This is merely a problem of calculating the number of ways to put $L-1$ partitioning bars (the cell boundaries) between $N$ balls with two additional fixed bars at the ends of the array (Chapter II.5 "Application to occupancy problems" in the textbook by Feller [88]). The answer is

$$D(N, n) = \binom{N + L - 1}{L - 1} \qquad (16.1)$$

The probability to have $k$ balls in a given box is proportional to the number of arrangement of $N-k$ balls in $L-1$ boxes.

$$P_k = \frac{D(N-k, L-1)}{D(N, L)} = \binom{N+L-k-2}{L-2} \div \binom{N+L-1}{L-1} \qquad (16.2)$$

In the limit of large $N$ and $L$, Eq.(16.2) has the asymptotic (Chapter II.11 "Problems and complements of a theoretical character" in Ref. [88]):

$$P_k \propto \frac{L}{N+L}\left(\frac{N}{N+L}\right)^k . \qquad (16.3)$$

The distribution is exponential, just like the distribution of the sizes of genera of the same age in Yule's model (Eq.(2.4)). There is a reason for that. We had quantified the arrangements of $N$ balls in $L$ bins. One way to practically obtain these arrangements is to first put the bars and then start adding balls between them. So when we add the first ball it can go to any interval between bars, that is into any bin, with equal probability. When we add the second ball, it will have a higher probability to get into the bin, which already has a ball, since it has two spaces between the ball, which is in it, and two boundaries. In general, when the bin has $k$ balls it has $k+1$ spaces and the probability for a new ball going into it is proportional to $k+1$. We can now derive Eq.(16.3) by Yule's method. Suppose that we start with $L$ empty bins. Each empty bin corresponds to the original specie of the genera. We add balls after equal time intervals and $N$ plays the role of time. If we have already $N$ balls in the system and $k$ balls in the bin of interest than the probability that the next ball goes to this bin is equal to $\frac{k+1}{N+L}$. This is equivalent to having time dependent rate of specific mutations. We can use Eq.(2.4), but should replace $st$ with $\int_0^N \frac{dx}{x+L} = \ln\left(\frac{N+L}{L}\right)$. We also should replace $k$ with $k+1$ since the bin plays the role of the initial specie. After substituting the above into Eq.(2.4) we obtain Eq.(16.3).

Apart from the large $L$ asymptotic, small $L$ cases are of interest. In the case $L=2$ we get Markov's urn model (Section 7). The case $L=3$ describes a Bose-Einstein condensate of spin-1 particles. Using Eq.(16.2) we get that the probability for $N_1$ bosons to be in $s_z = 1$ state is $P_{N_1} = 2\frac{N+1-N_1}{(N+1)\times(N+2)}$. In the limit of large $N$ the probability density of fraction of condensate in $s_z = 1$ state, $n_1 = N_1/N$, is $p(n_1) = 2(1 - n_1)$. This means that, for example, with 1% probability 90% or more of bosons will be in $s_z = 1$ state. That is the system shows spontaneous magnetization. The projection of magnetization on z-axis is $M_z = N_1 - N_{-1}$ and its distribution one can calculate similarly to how we got other results in this section. The result is that the probability density of the magnetization per spin is $p(m_z) = 1 - |m_z|$. This Bose-ferromagnetism is a result of cumulative advantage principle: bosons tend to transit into that quantum state where many of their boson buddies had already accumulated. The phenomenon of Bose ferromagnetism was predicted with the help of one of the authors of this paper [86]. As the reader surely understood, it was merely a re-discovery of what was known long before [87].

In 1977 Ijiri and Simon [22] had shown that if we modify the scheme in a way that we not only add balls but also add bins (quantum states) we get Simon's model from Bose-Einstein statistics. With probability $1-\alpha$ we just add a ball and its probability to go to a particular bin is proportional $k+1$, where $k$ is bin's occupation. With probability $\alpha$ we add a new bin. This leads exactly to Simon's process. This procedure was rediscovered in 2001 by Bianconi and Barabasi [23] who used the language of quantum states and Bose statistics. It was rediscovered again in 2003 by Chung, Handjani, and Jungreis [20], who used the language of bins and balls.

## 17. Branching process. Bienaymé-Cournot, Galton-Watson et al.

Textbooks (Harris [38], Athreya and Nay [79]) say that Galton and Watson discovered the branching process in $1870^{\text{th}}$. Therefore, it is often called the Galton-Watson process. However, Bru, Jongmans, and Seneta [54] reported that a mathematical solution of a branching process model appears in the 1847 book by Cournot [55]. He considered a gambler who pays an *écue* for a ticket which can win $0, 1,\ldots n$ *écues* with the probabilities $p(0)$, $p(1),\ldots p(n)$. The gambler at the beginning of the game has one *écue*. He buys a ticket and in the second round uses all the money (if any) that he won in the first round to buy new tickets. The game continues so that all the money, won in the preceding round, is used to buy tickets in the next round. Cournot asks the question: what is the probability that the gambler will eventually go bankrupt. Let us denote as $p_b^k$ the probability that he is bankrupt by the $k$th round. Then the probability that he will be bankrupt after $k+1$ rounds is equal to:

$$p_b^{k+1} = \sum_{n=0}^{\infty} p(n)\left(p_b^k\right)^n \tag{17.1}$$

The probability that he eventually goes bankrupt, $p_b$, is given by the obvious self-consistency equation

$$p_b = \sum_{n=0}^{\infty} p(n)\left(p_b\right)^n \tag{17.2}$$

Cournot explicitly solved the $n = 2$ case. Here the self-consistency equation becomes:

$$p(2)\left(p_b\right)^2 - \left(p(0) + p(2)\right)p_b + p(0) = 0 \tag{17.3}$$

Here we used the condition:

$$p(0) + p(1) + p(2) = 1 \tag{17.4}$$

There are two solutions to Eq. (17.3):

$$p_b' = 1 \; ; \; p_b'' = \frac{p(0)}{p(2)} \tag{17.5}$$

When $p(2) < p(0)$, $p_b'' > 1$ and the only relevant root is $p_b' = 1$. When $p(2) = p(0)$, the two roots coincide. It is also obvious that when $p(0) = 0$ the relevant root is $p_b'' = 0$. As $p_b$ should be a continuous function of $p(0)$ the only possibility is that for $p(2) > p(0)$ the relevant root is $p_b''$. The critical condition $p(2) = p(0)$ can be rewritten using Eq. (17.4) as

$$p(1) + 2p(2) = 1 \tag{17.6}$$

which means that the expectation value of the win is equal to the price of the ticket.

In a footnote, Cournot mentions that the gambler problem he solved is similar to the family problem of Bienaymé. In 1845 Bienaymé (the discovery of his research note was reported by Heyde and Seneta [53] and the note itself was reprinted by Kendall [73]) had considered the following problem. In each generation, $p(0)$ percent of the adult males have no sons, $p(1)$ have one son and so on. What is the probability that the family (or the family name) gets extinct? He wrote that there are two different regimes: one when the probability of extinction is unity and another when it is less than unity. The point where the average number of sons equals unity separates the two regimes. He did not supply any mathematical proof, however, and only wrote that the mathematical solution will be published later. Bru, Jongmans, and Seneta [54]believe that it indeed followed and that Cournot had read it.

In 1873 Francis Galton, who was upset by extinction of many prominent British families, published in the newspaper *Educational Times* the problem identical to that of Bienaymé. Watson proposed a solution to the problem and in 1875 together with Galton they published their paper [37]. Watson invented the method of generating functions, which he defined as:

$$f(z) = \sum_{n=0}^{\infty} p(n) z^n \ . \tag{17.7}$$

These functions have many useful properties, including that the generating function for the number of grandsons is $f_2(z) = f(f(z))$. To prove this, notice that if we start with two individuals instead of one, and both of them have offspring probabilities described by $f(z)$, their combined offspring has generating function $(f(z))^2$. This can be verified by observing that the $n$th term in the expansion of $(f(z))^2$ is equal to $\sum_{m=0}^{n} p(n-m) p(m)$, which is indeed the probability that the combined offspring of two people is $n$. Similarly one can show that the generating function of combined offspring of $n$ people is $(f(z))^n$. The generating function for the number of grandsons is thus:

$$f_2(z) = \sum_{n=0}^{\infty} p(n) (f(z))^n = f(f(z)).$$

In a similar way one can show that the generating function for the number of grand-grandsons is $f_3(z) = f_2(f(z))$ and in general:

$$f_{k+1}(z) = f_k(f(z)). \tag{17.8}$$

One can use the method of generating functions to find the probability of extinction of a family. Obviously, the probability to be extinct after $k$ generation is equal to $p_{ext}^k = f_k(0)$ and after $k+1$ generations it is equal to $p_{ext}^{k+1} = f_{k+1}(0)$. Using Equation (17.8) this can be rewritten as

$$p_{ext}^{k+1} = f(p_{ext}^k). \tag{17.9}$$

The probability of extinction after infinite number of generations, $p_{ext}$, is the fixed point of Equation (17.9):

$$p_{ext} = f(p_{ext}). \tag{17.10}$$

One obvious solution is $p_{ext} = 1$ (note that $f(1) = 1$, as it is the sum of all offspring probabilities). Watson [37] had found this solution and concluded that all families always get extinct.

In 1929 Danish mathematician Erlang was upset by the extinction of a prominent Danish family. He published in the journal *Matematisk Tidsskrift* a problem identical to that published by Galton 56 years earlier (see [49], [50]). Steffensen published the solution [51] in the same journal the next year[4]. He re-invented the generating function formalism, but, unlike Watson, correctly solved the problem of extinction. The fate of families depends on the average number of sons

---

[4] Christensen also solved the problem, but his solution was received by the journal few weeks after Steffensen's and for that reason not published at the time. It was only published in 1976 [52].

$\lambda = \sum n p(n) = \left[ f'(z) \right]_{z=1}$.

When $\lambda < 1$ we always have $f(0) = p(0) > 0$. The curve $f(z)$ intersects the line $y = z$ at the point $z = 1$. Since $f'(1) < 1$, $f(z)$ must be above the line $y = z$ when it approaches the intersection from the left (see Figure 3). Since $f(z)$ is a polynomial with all positive coefficients – it is convex up. Thus $f(z) > z$ in the interval $0 \le z < 1$. Therefore Eq.(17.9) has only one fixed point in the interval between zero and unity: $p_{ext} = 1$. This fixed point is stable. Clearly, if $p_{ext}^k = 1 - \varepsilon$, Eq.(17.9) gives $p_{ext}^{k+1} = 1 - \lambda \varepsilon$, which is closer to the fixed point when $\lambda < 1$. Accordingly, when $\lambda > 1$ the fixed point $p_{ext} = 1$ is unstable. It is also clear that in the case $\lambda > 1$, $f(z)$ must be below the line $y = z$ when it approaches the intersection at $z = 1$ from the left (see Figure 3). As $f(0) = p(0) \ge 0$ and $f(z)$ is convex up, there is a single intersection of $f(z)$ and $y = z$ in the interval $0 \le z < 1$. It is also clear that this fixed point is stable.

Thus when $\lambda < 1$, all families eventually extinct (this is called subcritical branching process). When $\lambda > 1$, some of the families get extinct, while others continue to exist forever (this is called supercritical branching process). The intermediate case, $\lambda = 1$, is critical branching process, where all families extinct, just like in a subcritical process, though some of them only after very long time.

These results were obtained by Steffensen [51], who is normally credited with being the first to solve the extinction problem correctly (see the textbook by Harris [38]). However, the extinction probability in family problem is equivalent to the bankruptcy probability in the gambler problem. As we have seen Cournot or Bienaymé [55] obtained the correct solution long before even Watson.

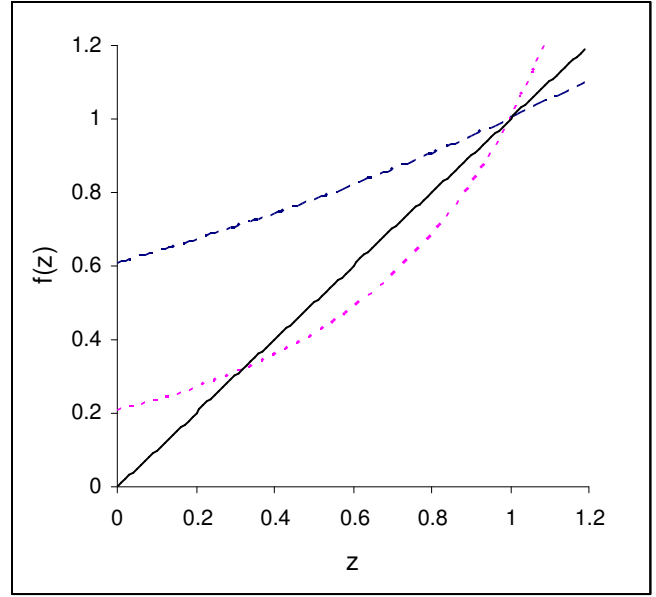

**Figure 3.** Generating functions of branching processes with Poisson offspring distributions. Dotted line is for $\lambda = 0.5$ and dashed line for $\lambda = 4$. Intersections of generating functions with the line $y = z$ correspond to solutions of Eq.(17.10).

In the period between Galton's and Erlang's reinventions of branching processes one more reinvention took place. In 1922 British biologist, Ronald A. Fisher, studied the problem of the spread of a mutant gene [56]. The problem is mathematically identical to that of Bienaymé and Galton. One just needs to replace "family name" with "gene" and the number of individuals in future generation with the number of genes. Fisher reinvented the generation functions method and the recursion relation (17.8). He then used it to compute numerically the survival probabilities after specific number of generations. In 1929 Fisher published the book [74] where he again described the results of his 1922 article [56]. Interestingly, he cited Galton in that book, but not the Galton-Watson process. The other Galton's work, "Hereditary Genius," which Fisher cited also concerned family extinction. However, that time Galton explained it not by an ordinary law of chances but by an introduction of an heiress into the family. His reasoning was that peers were likely to marry heiresses to supplement their dignity with monetary income. At the same time, an heiress who was the only child in the family is likely to be less fertile than a woman who has many brothers and sisters.

Branching process is similar to Yule's process. If we apply Yule's model to find the distribution of species by number of individuals than "specie" will be replaced with "individual" and "genus" with "specie". Specific mutation will correspond to a birth without mutation and generic mutation to specific mutation. These are only verbal changes, which do not affect math. Now remember that, unlike species, all individuals are mortal. We will have to adjust Yule's model for that, and this will transform it into Galton-Watson model.

The scientists will appreciate the following illustration of the connection between Yule-Simon and Galton-Watson processes. A model where scientist writing a manuscript picks up few random papers cites them, and copies a fraction of their references leads to Yule-Simon process. A modification of that model where scientist in the process of writing a manuscript picks up few random *recent* papers cites them, and copies a fraction of their references leads to Galton-Watson process. The only difference between the two models is the word *recent*. See the Mathematical theory of citing [40] for details.

## 18. Chemical chain reactions

In 1935, Semenoff [57] developed a theory of branching chains in chemical reactions. Let us illustrate it on the example of burning hydrogen. Consider a mixture of oxygen and hydrogen gases. When two molecules $O_2$ and $H_2$ collide, the reaction does not happen unless the temperature, and, correspondingly, kinetic energies of the molecules, is very high. This is because to dissociate an oxygen molecule one needs to overcome an energy barrier. However if we have a free oxygen atom then the following exothermic reaction will always proceed:

$$O + H_2 = H_2O^*$$

The energy released in the above reaction becomes vibrational energy of the water molecule. We denote by * that the molecule has excess energy. This energy by the way of collisions can be redistributed over many molecules, causing mere heating. Alternatively, the excited water molecule can dissociate an oxygen molecule:

$$H_2O^* + O_2 = H_2O + O + O$$

If the latter reaction happens, then instead of one oxygen atom we got two. If the probability of the latter reaction is more than one half, then we have a supercritical branching process. This probability depends on temperature, density and relative concentration of the gases. Thus, one can change it by changing those parameters. The supercritical chemical chain reaction is commonly known as "explosion."

Semenoff did not refer to any prior work on branching processes in his book. He understood the difference between subcritical and supercritical branching processes, but mathematically did not go any further than that.

## 19. Nuclear chain reactions

Nuclei of uranium can spontaneously fission, i.e. split into several smaller fragments. During this process, two or three neutrons are emitted. These neutrons can induce further fissions if they hit other uranium nucleuses. As the size of a nucleus is very small, neutrons have good chance of escaping the mass of uranium without hitting a nucleus. This chance decreases when the mass increases, as the probability of hitting a nucleus is proportional to the linear distance a neutron has to travel through uranium to escape. The fraction of neutrons that escape without producing further fissions is analogous to the fraction of the adult males who have no sons in Galton-Watson model. The neutrons produced in a fission induced by a particular neutron are analogous to sons. Critical branching process corresponds to a critical mass. A nuclear explosion is a supercritical branching process. It is not surprising that branching process was re-invented in this context. Hawkins and Ulam [61] did this in 1944. They re-invented the whole generating function method. They went a bit further then their predecessors, however. They studied the probability distribution, $P(n)$, of the total number of descendants, which is the sum of the numbers of sons, grandsons, grand-grandsons and so on (to be precise we include self in this sum just for mathematical convenience). They defined the corresponding generating function:

$$g(z) = \sum_{n=1}^{\infty} P(n)z^n . \qquad (19.1)$$

Using an obvious self-consistency condition (similar to the one in Eq.(17.10)) we get:

$$zf(g) = g \qquad (19.2)$$

This result was rediscovered in 1948 by Good [93]. He had referred to Galton and Watson, and pointed out that Fisher and Woodward (see the end of Section 20) had re-invented Galton-Watson process. He only did not know that his new result was also obtained before.

Otter [75] solved the above equation using Lagrange expansion (see Eq.(A3)) and got:

$$g = \sum_{n=1}^{\infty} \frac{z^n}{n!} \left[ \frac{d^{n-1}}{d\omega^{n-1}} (f(\omega))^n \right]_{\omega=0} . \qquad (19.3)$$

By comparing Eq.(19.1) and Eq.(19.3) we get:

$$P(n) = \frac{1}{n!} \left[ \frac{d^{n-1}}{d\omega^{n-1}} (f(\omega))^n \right]_{\omega=0} . \qquad (19.4)$$

## 20. Cascade electron multipliers

When certain surfaces are bombarded with electrons, they emit secondary electrons. The number of secondary electrons is proportional to the number of primary electrons, and the factor of proportionality may be as big as ten. Therefore, secondary emission can be used to amplify a small initial electron current. Since the amplification factor is not big, for practical applications, it was necessary to develop a multistage electron multiplier [89]. In such devices, the initial electron stream is impinged upon a target. The secondary electrons from this target are directed on to a second target, producing still further electrons, the multiplication being repeated many times. This is, of course, a branching process with secondary electrons equivalent to children and stages to generations.

In 1938 Shockley and Pierce [90] developed the theory of noise in such devices. Suppose that we have many identical stages. Each of them produces for one primary electron an average of $\lambda$ secondary electrons and the standard deviation of the number of secondary electrons is $\sigma$. Suppose that the number of electrons, $n_k$, after $k$th stage is described by the probability distribution $p_k(n_k)$ with the average $m_k$ and standard deviation $\sigma_k$. After the next stage we obviously have

$$m_{k+1} = \lambda m_k \qquad (20.1)$$

and

$$\sigma_{k+1}^2 = \left\langle \left( p_{k+1}(n_{k+1}) - \lambda m_k \right)^2 \right\rangle = \sum_{l=0}^{\infty} p_k(l) \left\langle \left( p_{k+1}(n_{k+1} \mid n_k = l) - \lambda m_k \right)^2 \right\rangle \qquad (20.2)$$

The expectation value in Eq. (20.2) can be rewritten as:

$$\left\langle \left( p_{k+1}(n_{k+1} \mid n_k = l) - \lambda m_k \right)^2 \right\rangle = \left\langle \left( p_{k+1}(n_{k+1} \mid n_k = l) - \lambda l + \lambda l - \lambda m_k \right)^2 \right\rangle =$$
$$\left\langle \left( p_{k+1}(n_{k+1} \mid n_k = l) - \lambda l \right)^2 \right\rangle + \left\langle \left( p_{k+1}(n_{k+1} \mid n_k = l) - \lambda l \right) \right\rangle (\lambda l - \lambda m_k) + (\lambda l - \lambda m_k)^2 \qquad (20.3)$$

The first term in Eq.(20.3) equals

$$\left\langle \left( p_{k+1}(n_{k+1} \mid n_k = l) - \lambda l \right)^2 \right\rangle = l \left\langle \left( p_{k+1}(n_{k+1} \mid n_k = 1) - \lambda \right)^2 \right\rangle = l \sigma^2$$

and the second equals zero. After substituting this into Eq.(20.2) we get

$$\sigma_{k+1}^2 = \sigma^2 \sum_{l=0}^{\infty} p_k(l) l + \lambda^2 \sum_{l=0}^{\infty} p_k(l)(l - m_k)^2 = \qquad (20.4)$$

$$m_k \sigma^2 + \lambda^2 \sigma_k^2$$

This recursion relation can be iterated to get

$$\sigma_k^2 = \lambda^{2k} \sigma_0^2 + m_0 \sigma^2 \frac{\lambda^{2k} - \lambda^k}{\lambda(\lambda - 1)} \qquad (20.5)$$

Here $m_0$ and $\sigma_0$ describe primary current. The first term in Eq.(20.5) is just amplified noise in primary current. The second term describes additional noise introduced by the device.

These results had been already in 1933 obtained by Steffensen [91] who used generating function formalism. By doubly differentiating Eq.(17.8) we get

$$f_{k+1}''(1) = [f'(1)]^2 f_k''(1) + f_k'(1) f''(1)$$
$$= \lambda^2 f_k''(1) + m_k f''(1) \qquad (20.6)$$

Note that $f''(1) = \sum_{n=1}^{\infty} n(n-1) p(n) = \sigma^2 - \lambda + \lambda^2$ . A similar relation holds for $f_k''(1)$. By substituting this into Eq. (20.6) we recover Eq.(20.4).

In 1947 Woodward [92] developed a mathematical theory of cascade multiplication. He re-invented generating function method, derived Eq.(20.5) the same way as Steffensen did, and correctly solved the zero output problem (which is analogous to family extinction).

## 21. Molecular size distribution in polymers

In 1941, branching process was re-invented by Flory who studied the formation of polymers. He considered what he called trifunctional units, schematically shown in the upper left corner of Figure 4. Each such monomer unit consists of a node with three functional units attached to it. Each of three functional units can form a chemical bond with any functional unit of another trifunctional unit. This way the polymer molecules, like the one shown in Figure 4, are built. Flory considered the following problem. Suppose the fraction of functional units, which had reacted (connected with other functional units), is $\alpha$. What is the distribution of molecules by weight? We shall reproduce the solution given by Flory (apart from correcting his errors).

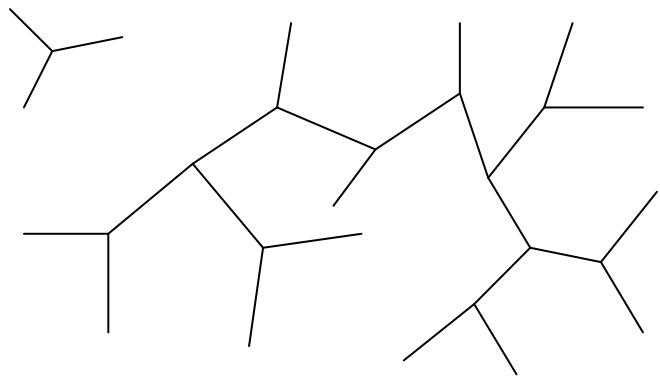

**Figure 4.** In the upper left corner we show a trifunctional monomer. The rest of the picture shows a three-dimensional polymer molecule formed by such monomers.

Let us select a monomer at random. What is the probability that it is a part of an $n$-monomer molecule? Let us select one of its three functional units at random. At the outer end of this unit, there may be another monomer connected to it, or there may be none. However, at the inner end there is always a branch of two functional units belonging to the same monomer. We first compute the probability distribution of the weight of the part of the molecule hanging on those two branches. It is easy to do under the approximation that there are no intermolecular reactions. That is two functional units of a two monomers already belonging to one polymer molecule cannot react and form a bond with each other. In graphical representation of the molecule, this condition corresponds to allowing no loops. The following recursion relation gives the probability $\overline{W}_n$ that that part of a molecule consists of $n$ monomers:

$$\overline{W}_n = \alpha^2 \sum_{k=0}^{n-1} \overline{W}_{n-k-1} \overline{W}_k \qquad (21.1)$$

where we for convenience defined $\overline{W}_0 = \dfrac{1-\alpha}{\alpha}$ .
Flory could not solve the recursion and William C. Taylor whom he acknowledges in footnote 7 did it for him.

The solution, of course, uses a generating function

$$g(z) = \sum_{n=0}^{\infty} \overline{W}_n z^n \ . \qquad (21.2)$$

One can see that

$$\left(g(z)\right)^2 = \sum_{k=0}^{\infty} \sum_{l=0}^{\infty} \overline{W}_k z^k \overline{W}_l z^l = \sum_{n=0}^{\infty} z^n \sum_{k=0}^{n} \overline{W}_k \overline{W}_{n-k}$$

Substituting Eq.(21.1) into the above equation we get

$$\left(g(z)\right)^2 = \frac{1}{\alpha^2} \sum_{n=0}^{\infty} \overline{W}_{n+1} z^n = \frac{1}{\alpha^2 z} \left(g(z) - \overline{W}_0\right)$$

which is a quadratic equation for $g(z)$ . The solution of it is

$$g(z) = \frac{1}{2\alpha^2 z} \left(1 - \sqrt{1 - 4\alpha(1-\alpha)z}\right) \qquad (21.3)$$

We eliminated the alternative plus sign because it gives negative probabilities. After applying binomial expansion to the square root and using the definition of the generating function in Eq.(21.2) we get

$$\overline{W}_n = \frac{1}{\alpha^2} \frac{\left(\alpha(1-\alpha)\right)^{n+1}(2n)!}{n!(n+1)!} \qquad (21.4)$$

Remember that $\overline{W}_n$ is the probability that the part of the molecule consisting of a randomly selected trifunctional unit and everything connected to its two functional consists of $n$ monomers. There is also a third functional unit. With probability $1-\alpha$ nothing is connected to that unit. With probability $\alpha$ another trifunctional monomer is connected to it. The probability that that other monomer and everything attached to its other two functional units comprise together $k$ monomers is equal to $\overline{W}_k$ . The probability $W_n$ that the total number of monomers in a molecule, of which a randomly selected monomer is a part, equals $n$ is:

$$W_n = (1-\alpha)\overline{W}_n + \alpha \sum_{k=1}^{n-1} \overline{W}_{n-k} \overline{W}_k$$

This after some transformations can be reduced to

$$W_n = \frac{1}{\alpha} \left(\overline{W}_{n+1} - \alpha(1-\alpha)\overline{W}_n\right).$$

After substituting Eq. (21.4) into the above equation we get

$$W_n = \frac{3\alpha^{n-1}(1-\alpha)^{n+2}(2n)!}{(n-1)!(n+2)!} \qquad (21.5)$$

When $n$ is large we can use Stirling formula for the factorials and get

$$W_n \simeq \frac{(1-\alpha)^2}{\sqrt{\pi}\alpha} \frac{\left(4\alpha(1-\alpha)\right)^n}{n^{3/2}}$$

This is a pure power law when $\alpha = 1/2$ and a power law with an exponential cutoff for all other values of $\alpha$.

$W_n$ is the probability that randomly selected monomer belongs to molecule consisting of $n$-monomers. It can also be interpreted as the weight fraction of the $n$-mers in the solution. The number of $n$-mers in the solution is therefore:

$$m_n = NW_n/n \ , \qquad (21.6)$$

where $N$ is the total number of monomers in the system.

Let us compute the sum of the masses of all polymer molecules of finite sizes. It is equal to

$$\sum_{n=1}^{\infty} W_n = \frac{1}{\alpha}\left(\sum_{n=1}^{\infty} \overline{W}_{n+1} - \alpha(1-\alpha)\sum_{n=1}^{\infty} \overline{W}_n\right) =$$

$$\frac{1}{\alpha}\left(\sum_{n=0}^{\infty} \overline{W}_n - \overline{W}_1 - \overline{W}_0 - \alpha(1-\alpha)\left(\sum_{n=0}^{\infty} \overline{W}_n - \overline{W}_0\right)\right)$$

After observing that $\sum_{n=0}^{\infty} \overline{W}_n = g(1)$ and doing some algebra we get

$$\sum_{n=1}^{\infty} W_n = \begin{matrix} 1 & when\ \alpha \le 1/2 \\ ((1-\alpha)/\alpha)^3 & when\ \alpha > 1/2 \end{matrix}$$

Therefore, when $\alpha$ exceeds the critical value $\alpha_c = 1/2$ some mass appears to be lost. It is because of the formation of one giant molecule, which absorbed a finite fraction of all monomers in the system. This is a phase transition called gelation.

This derivation can be formulated in terms of Galton-Watson model. The only complication is that the first selected at random monomers can have up to three sons and all of his descendants only up to two. The size of the molecule corresponds to the total number of descendants. The sum of the masses of finite molecules corresponds to the probability of family extinction. To illustrate this we will redo the calculation for a general case of a polymer formed of $q$-functional monomers. As one of the functional units connects the monomer to its father, we should study the branching process with up to $q-1$ sons, each of which can be born with the probability $\alpha$. The generating function for the offspring probability is

$$f(z) = \sum_{k=0}^{q-1} \binom{q-1}{k}(1-\alpha)^{q-1-k}\alpha^k z^k$$

$$= (1-\alpha + \alpha \times z)^{q-1} \tag{21.7}$$

The generating function for the total number of descendants satisfies the self-consistency relation of Eq.(19.2). Substituting in it Eq.(21.7) we get:

$$zg(z) = (1-\alpha + \alpha \times g(z))^{q-1}. \tag{21.8}$$

Now let us find the total offspring of a randomly selected monomer. This one does not have a father and, thus, can have up to $q$ sons. The generating function for its total number of descendants is

$$\phi(z) = z(1-\alpha + \alpha \times g(z))^q$$

Using Eq. (21.8), we can rewrite this as:

$$\phi(z) = g(z)(1-\alpha + \alpha \times g(z)) \tag{21.9}$$

Substituting Eqs (21.8) and (21.9) into Eq. (A2) we get the following equation for the total number of descendants probabilities (or, equivalently, of the weight fraction of $n$-mers):

$$W_n = \frac{1}{2\pi \times i} \times$$

$$\oint dg\, \frac{(1-\alpha + \alpha \times g)^{n(q-1)}}{ng^n}((1-\alpha) + 2\alpha \times g) =$$

$$\left(\binom{n(q-1)}{n-1} + 2\binom{n(q-1)}{n-2}\right)\frac{(1-\alpha)^{n(q-2)+2}\alpha^{n-1}}{n}$$

The above equation can be reduced to

$$W_n = \frac{q(1-\alpha)^2}{\alpha}\frac{\left(\alpha(1-\alpha)^{q-2}\right)^n (qn-n)!}{(n-1)!(qn-2n+2)!} \tag{21.10}$$

This formula was first derived by Stockmayer [60], who solved the problem using a complicated combinatorial method. In the case $q = 3$ Eq.(21.10) reduces to Eq.(21.5).

By applying Stirling's formula to Eq.(21.10) we get:

$$W_n \approx \frac{(1-\alpha)^2}{\sqrt{2\pi\alpha}}\sqrt{\frac{q-1}{(q-2)^5}}\times\frac{x^n}{n^{3/2}}.\qquad (21.11)$$

Here

$$x = \alpha(1-\alpha)^{q-2}\frac{(q-1)^{q-1}}{(q-2)^{q-2}}\qquad (21.12)$$

One can show that $x(\alpha)$ reaches maximum when

$$\alpha = \frac{1}{q-1}\equiv\alpha_c\qquad (21.13)$$

and that $x(\alpha_c)=1$. Thus, when $\alpha=\alpha_c$, $W_n$ decays as a power law and when $\alpha\neq\alpha_c$, $W_n$ decays exponentially with $n$. When $|\alpha-\alpha_c|<1$, Eq. (21.12) can be expanded as

$$x = 1 - \frac{(q-1)^3}{2(q-2)}(\alpha-\alpha_c)^2$$

After substituting this into Eq. (21.11) we get

$$W_n \sim \frac{1}{n^{3/2}}e^{-n/n_c}\qquad (21.14)$$

where

$$n_c \sim (\alpha-\alpha_c)^{-2}\qquad (21.15)$$

The weight fraction of finite molecules we can find using extinction probability. For the case of the branching process with up to $q-1$ sons, each of which can be born with the probability $\alpha$, the extinction probability satisfies the following self-consistency equation:

$$\bar{p}_{ext} = \left(1-\alpha+\alpha\times\bar{p}_{ext}\right)^{q-1}.\qquad (21.16)$$

The probability of extinction of the family of a randomly selected monomer, which does not have a father and, thus, can have up to $q$ sons is:

$$p_{ext} = \left(1-\alpha+\alpha\times\bar{p}_{ext}\right)^q$$

Using Eq. (21.16) it can be rewritten as

$$p_{ext} = \bar{p}_{ext}\left(1-\alpha+\alpha\times\bar{p}_{ext}\right)$$

Equation (21.16) can be easily solved for $q\leq 3$ but becomes complicated for bigger $q$. It is, however, easy to find its asymptotic for the case $\alpha-\alpha_c<<1$. We get

$$p_{ext} \approx 1 - \frac{2(1+\alpha_c)}{\alpha_c^2(q-2)}(\alpha-\alpha_c)$$

The mass fraction of the giant molecule is thus

$$W_g = 1 - p_{ext} \approx \frac{2(1+\alpha_c)}{\alpha_c^2(q-2)}(\alpha-\alpha_c).\qquad (21.17)$$

## 22. Percolation Process

In 1957 Broadbent and Hammersley [94] published their paper which founded percolation theory. This was not a re-invention since they referred to the paper by Good [93], who cited Galton, Watson and Steffensen. We will discuss percolation, however, to have a broader review of the fields of application of branching processes. Broadbent and Hammersley considered a system of channels leading from the original ancestor such that each channel divides into precisely two channels at each stage. Each of these channels has, independently of the other channels, a probability $1-p$ of being dammed. The question they asked was what is the probability, $p_f$, that only finite number of channels will be flooded. The problem is similar to the gambler problem discussed in the beginning of Section 17, and $p_f$ is equal to bankruptcy probability. As the average number of not dammed descendant channels is $2p$, the critical value is $p_c = 1/2$. When $p < p_c$, $p_f = 1$. When $p > p_c$, $p_f$ is equal to $p_b''$ in Eq.(17.5). By substituting $p_0 = (1-p)^2$ and $p_2 = p^2$ into Eq.(17.5) we get $p_f = (1-p)^2/p^2$.

Percolation problem has even greater similarity to Flory's polymer problem discussed in Section 21. The fraction of dammed channels corresponds to the fraction of unreacted bonds. This is why not only the general methods of solution, but offspring distributions, generating functions and all the mathematical results are identical for the two problems. For example if we replace in Eq.(21.7) the fraction of reacted bonds $\alpha$ with the fraction of not dammed channels $p$ and $q-1$ with the number of descendant channels we will get the generating function for the channel problem.

Let us now consider the channel configuration shown in Figure 1a in Section 10 and use the theory of branching processes to solve the same percolation problem we already studied using Renormalization Group. The problem was formulated in terms of removing nodes (removing a node is equivalent to damming all the channels leading to it). It is, thus, more convenient to speak in terms of nodes. Let us select a random node. It is connected to six neighboring nodes, which we will call its children. Each of the children is connected to additional five nodes, which we will call their children. Here we encounter a difficulty. There are instances when say nodes $B$ and $C$ are both children of node A, but in addition $B$ is a child of $C$ and $C$ is a child of $B$. This is why all results obtained using methods of the theory of branching processes will be only approximate. When we neglect the effect we just described the problem becomes identical to Flory's problem with $q=6$. Now Eq.(21.13) gives $p_c = 1/5$. This is way off the exact result $p_c = 1/2$ and shows the importance of the effect we had neglected.

The weight fraction of $n$-mers corresponds to the weight fraction of connected clusters of size $n$. Thus the number of nodes in the largest connected cluster is given by Eq.(21.15). Since the lattice in question is two dimensional it is reasonable to suppose that the size of the cluster is proportional to the square root of the number of nodes in it. Thus, the size of the largest connected cluster is equal to

$$\xi(p) \sim (p - p_c)^{-1}.$$

Though we do not get exact percolation results using theory of branching processes, we still get qualitatively correct behavior.

In his 1985 and now classic textbook on percolation [95], Stauffer writes that Flory developed percolation theory before Broadbent and Hammersley. This is of course true. It is also true that Broadbent and Hammersley were not familiar with the work of Flory and did not cite it. However, Broadbent and Hammersley did cite the work of Galton and Watson. Stauffer in his turn did not mention branching process in his book. In 2000, a group of scientists that included Stauffer introduced the idea of social percolation [96]. Remarkably, the work of Galton and Watson could be classed as such: it studied percolation of family names. The problem studied in [96] was a bit different: "a percolation phenomenon across the social network of customers". However, this problem had long ago been studied using branching processes. See the 1967 review of the epidemic of papers on "propagation of ideas, rumours and consumers' goods" [97].

## 23. Erdos-Renyi random graph

In their 1960 paper, Erdos and Renyi [62] considered $N$ vertices connected by $M$ random edges, such that any of $\binom{N}{2}$ possible edges was selected with equal probability. They were interested in properties of such graph in the limit $N, M \to \infty$. One question to ask about such graph is what is the degree distribution, or probability that a given vertex has $k$ edges. As each edge connects two vertices, the probability that a given edge connected to a given vertex is $2/N$. As there is total of $M$ edges then the average degree is $\lambda = 2M/N$. We have a large number of attempts to connect an edge to a given vertex and in each attempt the probability to connect is small. This satisfies the requirements of the Poisson process and thus we should get a Poisson degree distribution:

$$p(k) = \frac{\lambda^k}{k!}\, e^{-\lambda}.$$

Another thing they were interested in is the size distribution of the connected components. There are $\binom{N}{n}$ ways to select $n$ vertices out of $N$ total. According to Cayley's formula, there are $n^{n-2}$ different trees, which can be formed of $n$ vertices. The probability that a particular isolated tree will be realized is equal to the product of the probability of realization of $n{-}1$ of its edges and probability that $n$ of its vertices are not connected to outside vertices. The first is equal to $\left( M \middle/ \binom{N}{2} \right)^{n-1}$.

And the second to $\left( \left( 1 - M \middle/ \binom{N}{2} \right)^{N-n} \right)^n$. Thus, the expectation of the number of connected clusters of size $n$ is equal to

$$m_n = \binom{N}{n} \times n^{n-2} \times \left( M \middle/ \binom{N}{2} \right)^{n-1} \times$$

$$\left( \left( 1 - M \middle/ \binom{N}{2} \right)^{N-n} \right)^n$$

By taking the limit $N \to \infty$, keeping $M = N\lambda/2$, and using the definition of the exponent we get:

$$m_n = N \frac{n^{n-2} \lambda^{n-1}}{n!} e^{-\lambda n}. \tag{23.1}$$

Next Erdos and Renyi calculated the fraction, $W_f$ of vertices that belong to finite components:

$$W_f = \frac{1}{N} \sum_{n=1}^{\infty} n m_n \tag{23.2}$$

After substituting Eq.(23.1) into Eq.(23.2) we get

$$W_f = \sum_{n=1}^{\infty} \frac{n^{n-1} \lambda^{n-1}}{n!} e^{-\lambda n} = \frac{1}{\lambda} \sum_{n=1}^{\infty} \frac{n^{n-1}}{n!} \left( \lambda e^{-\lambda} \right)^n \tag{23.3}$$

Next Erdos and Renyi use what they call a well known fact (the reader can verify it using the Lagrange expansion which is described in the Appendix) that the inverse function of

$$y = x e^{-x} \tag{23.4}$$

is

$$x = \sum_{n=1}^{\infty} \frac{n^{n-1}}{n!} y^n \tag{23.5}$$

The function $y(x)$, given by Eq. (23.4), equals zero when $x = 0$ and when $x = \infty$. It has a maximum when $x = 1$. Thus the inverse function, $x(y)$, has two branches: $0 \le x \le 1$ and $1 < x$. Eq. (23.5) gives us the first branch. After noticing the resemblance between Eq.(23.3) and Eq.(23.5) with $y = \lambda e^{-\lambda}$ we get

$$W_f = \frac{x'}{\lambda} \tag{23.6}$$

where $x'$ is the solution of the equation

$$x e^{-x} = \lambda e^{-\lambda} \tag{23.7}$$

satisfying $0 \le x \le 1$. When $\lambda \le 1$ the right solution of Eq.(23.7) is $x = \lambda$ and Eq. (23.6) gives $W_f = 1$. When $\lambda > 1$ the right solution of Eq.(23.7) is $x = x'$ and, since $x' < 1$, Eq. (23.6) gives $W_f < 1$. Thus, not all vertices belong to finite components. This means that the graph has a giant connected component. This is analogous to the gel molecule discussed in Section 21.

The above results can be derived far more straightforwardly using theory of branching processes. We select a vertex at random and calculate the probability $W_n$ that it belongs to a connected cluster of size $n$. As the offspring probability distribution is Poisson, the generating function is

$$f(z) = e^{(z-1)\lambda} \tag{23.8}$$

After substituting this into Eq. (19.4) we get

$$W_n = \frac{1}{n!} \frac{d^{n-1}}{dg^{n-1}} e^{n(g-1)\lambda} \bigg|_{g=0} = \frac{(n\lambda)^{n-1}}{n!} e^{-\lambda n} \tag{23.9}$$

By substituting the above equation into Eq. (21.6) we get Eq. (23.1). Note that unlike in the polymer case we did not have the problem of the first selected vertex having more offspring than the next. This is because the number of possible connections is infinite and probability of each connection is infinitely small. One can also obtain Eq. (23.9) by taking the limit $q \to \infty$ in Eq. (21.10).

After applying Stirling's formula to Eq. (23.9), we get the large $n$ asymptotic of $W_n$:

$$W_n \propto \frac{1}{\lambda \sqrt{2\pi} n^{3/2}} e^{-(\lambda - 1 - \ln \lambda)n}$$

The expression $\lambda - 1 - \ln \lambda$ reaches its minimum value of 0 when $\lambda = 1$. In this case $W_n$ follows a power law. When $\lambda \neq 1$, $W_n$ decays exponentially with $n$. When $1 - \lambda \ll 1$ the factor in the exponent can be approximated as: $\lambda - 1 - \ln \lambda \approx (1 - \lambda)^2 / 2$. Thus the exponential cut-off of the power law happens at $n_c \approx 2/(1 - \lambda)^2$.

The size of the giant connected component of the graph can also be computed using theory of branching processes. The probability that a randomly selected vertex belongs to a finite component equals extinction probability. The latter is the solution of the Eq.(17.9). After substituting Eq.(23.8) into Eq.(17.10) we get

$$p_{ext} = e^{(p_{ext} - 1)\lambda} \tag{23.10}$$

The above equation is equivalent to Eq.(23.7) with $x$ replaced with $\lambda p_{ext}$. It is impossible to solve exactly Eq.(23.10) but it is easy to compute the first term of its expansion in powers of $\lambda - 1$:

$$p_{ext} \approx 1 - 2(\lambda - 1).$$

The size of the giant connected component, $W_g$, equals $1 - W_f = 1 - p_{ext}$. Thus when $\lambda - 1 \ll 1$ we have:

$$W_g \approx 2(\lambda - 1). \tag{23.11}$$

The above equation can be obtained by substituting $\alpha = \lambda/q$ and taking the limit $q \to \infty$ in Eq. (21.17).

It is difficult to say for sure who was the first to point out the connection between the Erdos-Renyi random graph and the Galton-Watson branching process. Google book search for webpages containing both of the phrases "random graph" and "branching process" lead us to the book by R. Durrett [76] where this connection was discussed at length. Albeit no mention of who was the first to use this connection could be found in that book. To our email, Prof. Richard Durrett replied that he learned of that connection from Prof. Harry Kesten. The story turned out to resemble Milgram's six degrees of separation experiment on the graph of human social contacts [81]. Prof. Kesten in his turn sent us to Prof. David Aldous, who replied that, the earliest paper he knows which makes an explicit connection, is the 1990 paper by R.M. Karp [77]. Prof. Karp replied that he might be the first to apply the branching process to random graph. The chain has thus ended. However, another Google search, for webpages containing all of the four words (random, graph, branching, process) produced the 1985 article by John L. Spouge, "Polymers and Random Graphs: Asymptotic Equivalence to Branching Processes" [78]. This, probably, does not mean that it is the first article to point out the connection, but probably does mean that trying out different Google search phrases is more efficient than asking experts.

## 24. Smoluchowski coagulation

In 1916, Smoluchowski [63] considered a colloid suspension of particles, which collide with each other and stick together and resulting clusters in turn collide with each other forming larger clusters. The problem is easier to formulate mathematically when initially colloid consists of identical particles. Then the size of the cluster is simply described by the number of particles it consists of. The dynamics of cluster size distribution is described by the following rate equations (here $m_n$ is the number of clusters of size $n$):

$$\frac{dm_n}{dt} = \frac{1}{2}\sum_{k=1}^{n-1} K(k, n-k) m_k m_{n-k}$$
$$- m_n \sum_{k=1}^{\infty} K(n,k) m_k \qquad (24.1)$$

Smoluchowski considered the simplest case when the coagulation rate does not depend on the sizes of the particles and therefore the coagulation kernel is constant

$$K(n,k) = K.$$

With such kernel Eq.(24.1) becomes

$$\frac{dm_n}{dt} = \frac{K}{2}\sum_{k=1}^{n-1} m_k m_{n-k} - m_n K \sum_{k=1}^{\infty} m_k \qquad (24.2)$$

Let us first compute the total number of the clusters in the system

$$M = \sum_{n=1}^{\infty} m_n. \qquad (24.3)$$

After summing Eq.(24.2) for all $n$ we get:

$$\frac{dM}{dt} = -\frac{K}{2} M^2,$$

which has the solution

$$M = \frac{N}{1+\beta \times t}.$$

Here N is the initial number of particles in the system and $\beta = NK/2$. Let us recursively find the solution of Eq. (24.2) of starting with $m_1$. For the latter we have the following equation:

$$\frac{dm_1}{dt} = -m_1 KM = -m_1 \frac{2\beta}{1+\beta \times t}.$$

This has the solution

$$m_1 = \frac{N}{(1+\beta \times t)^2}.$$

Now we can substitute $m_1$ into the differential equation for $m_2$. After solving it (with the initial condition $m_2(0) = 0$) we get

$$m_2 = \frac{N\beta \times t}{(1+\beta \times t)^3}.$$

And in general

$$m_n = \frac{N(\beta \times t)^{n-1}}{(1+\beta \times t)^{n+1}}.$$

In 1962, McLeod [64] studied a more complicated case, where the kernel is

$$K(n,k) = K \times nk$$

After we substitute the above kernel into Eq.(24.1) the rate equations become

$$\frac{dm_n}{dt} = \frac{K}{2}\sum_{k=1}^{n-1} nk \times m_k m_{n-k} - n \times m_n \times K \sum_{k=1}^{\infty} k \times m_k$$

Obviously,

$$\sum_{k=1}^{\infty} k \times m_k = N, \qquad (24.4)$$

thus we have

$$\frac{dm_n}{dt} = \frac{K}{2}\sum_{k=1}^{n-1} nk \times m_k m_{n-k} - n \times m_n \times KN. \qquad (24.5)$$

One can try to solve this equation recursively, similar to how we did the constant kernel case. We get: $m_1 = N \exp(-\beta \times t)$, where $\beta = KN$. Further, we obtain $m_2 = N \dfrac{\beta \times t}{2} e^{-2\beta t}$ and $m_3 = N \dfrac{(\beta \times t)^2}{2} e^{-3\beta t}$.

It is easy to see by induction that in general

$$m_n = N(\beta \times t)^{n-1} \exp(-n\beta \times t) C_n, \qquad (24.6)$$

where $C_n$ is a time-independent coefficient.
Substituting Eq.(24.6) into Eq.(24.5) we get the following recursion relation for $C_n$:

$$(n-1)C_n = \frac{1}{2} \sum_{k=1}^{n-1} k C_k (n-k) C_{n-k} \qquad (24.7)$$

Next, following Ben-Naim and Krapivsky [65] we introduce the generating function:

$$g = \sum_{n=1}^{\infty} n C_n z^n \qquad (24.8)$$

After summing Eq.(24.7) over $n$ and using the definition of $g$ in Eq.(24.8) we get

$$g - \int \frac{g}{z} dz = \frac{1}{2} g^2 .$$

After differentiating we get $g' - \dfrac{g}{z} = gg'$ or

$$\frac{dg}{g}(1-g) = \frac{dz}{z} .$$

Taking into account that $g(0) = 0$ the solution of the above equation is:

$$g \exp(-g) = z \qquad (24.9)$$

Substituting Eq.(24.9) into Eq A2 and using the definition of $g$ in Eq.(24.8) we get

$$nC_n = \frac{1}{2\pi \times i} \oint dg \frac{\exp(-ng)}{ng^n} = \frac{n^{n-1}}{n!}$$

substituting this into Eq.(24.6) we get

$$m_n = N(\beta \times t)^{n-1} \frac{n^{n-2}}{n!} \exp(-n\beta \times t). \qquad (24.10)$$

This is exactly Erdos-Renyi formula Eq.(23.1) with $\lambda = \beta \times t$. The reason for this is that when we dynamically add edges to the graph the component size distribution is described by Smoluchowski equation. This was pointed out by Ben-Naim and Krapivsky [65]. Note that Eq.(24.10) is valid for all values of $\beta \times t$ and not only for $\beta \times t \leq 1$ as is claimed in some papers. Although for the case $\beta \times t > 1$, Eq.(24.4) no longer holds Eq.(24.5) is still correct, because the last term correctly describes the reduction in $m_n$ due to coagulation of clusters of size $n$ with clusters of all finite sizes and with the giant cluster.

## 25. Forced Smoluchowski kinetics

Another possibility to consider is when there is an influx of particles in the system.

$$\frac{dm_n}{dt} = \frac{K}{2} \sum_{k=1}^{n-1} m_k m_{n-k} - m_n K \sum_{k=1}^{\infty} m_k + S \times \delta_{1,n} \qquad (25.1)$$

By summing Eq. (25.1) over all $n$, we get for the total number of clusters (defined by Eq.(24.3)) the following equation:

$$\frac{dM}{dt} = -\frac{K}{2} M^2 + S \qquad (25.2)$$

It has the stable stationary solution

$$M = \sqrt{2S/K} . \qquad (25.3)$$

For the number of particles of size 1 we get the following equation:

$$\frac{dm_1}{dt} = -m_1\sqrt{2SK} + S \qquad (25.4)$$

This has a stationary solution

$$m_1 = \sqrt{S/(2K)}. \qquad (25.5)$$

For the stationary solution for $n \geq 2$ we get the following recursion relation:

$$m_n = \sqrt{\frac{K}{8S}} \sum_{k=1}^{n-1} m_k m_{n-k}. \qquad (25.6)$$

We define the generating function

$$g = \sum_{n=1}^{\infty} m_n z^n \qquad (25.7)$$

And after multiplying Eq.(25.6) by $z^n$, summing it over $n$ from 2 to infinity, and utilizing Eq. (25.7) we get:

$$g - m_1 z = \sqrt{\frac{K}{8S}} g^2 \qquad (25.8)$$

From Eqs. (25.8) and (25.5) we get:

$$g = \sqrt{\frac{2S}{K}} \left(1 - \sqrt{1-z}\right) \qquad (25.9)$$

Expanding the square root and using the definition of $g$ in Eq.(25.7) we get:

$$m_n = \sqrt{\frac{2S}{K}} \frac{(2n)!}{(2n-1)(n!)^2 4^n} \qquad (25.10)$$

Using Stirling' formula we get the large $n$ asymptotic:

$$m_n \approx \sqrt{\frac{S}{2\pi K}} \frac{1}{n^{3/2}}. \qquad (25.11)$$

This formula was, probably, first derived in 1972 by Klett [68]. A numerical solution of recurrence equations was published in 1965 by Quon and Mockros [69]. They inferred the 3/2 exponent by fitting the data.

The just described forced Smoluchowski kinetics was originally used to describe colloids, aerosols and similar things. However, recently physicists applied it to social phenomena. For example, Pushkin and Aref [80] used it to describe bank merging. While Pushkin and Aref did refer to and used the prior research on the subject, other physicists managed to re-invent Smoluchowski kinetics. Recently Kim *et al* [70] did this in the context of the science of networks (see Section 15 of this article). They proposed a model of a network evolving by merging and regeneration of nodes. They reported numerical simulations indicating that such dynamics leads to a power-law distribution of connectivity. However, they reported no analytical solution. One of the models that they considered is as follows. At each time step, two arbitrary selected nodes are merged and a new node is born, which connects to one arbitrary selected node. The problem is very similar to the one we just solved with the only difference is that instead of a size of coagulated particle, we have a connectivity of a node. The dynamics of the model is described by the following equations ( $P(k)$ is the probability distribution of connectivity, $k$) :

$$\frac{dP(k)}{dt} = \sum_{n=0}^{k-1} P(n) \times P(k-n) + P(k-1) - 3 \times P(k) + \delta_{k,1} \qquad (25.12)$$

The convolution describes influx of nodes into state $k$ by merging all possible pairs whose degrees add up to $k$. The second term describes influx into state $k$ resulting from a newborn node connecting to node of degree $k$-1. The third terms accounts for three possibilities to leave the state $k$: the node is one of two nodes selected for merging or the node is the one to which a newborn node connects. Note that there are no nodes of degree 0, because each newborn node is of degree 1 and afterwards degree only grows. Thus: $P(0) = 0$.

The stationary solution of the Equation (25.12) is:

$$3 \times P(k) =$$
$$\sum_{n=0}^{k-1} P(n) \times P(k-n) + P(k-1) + \delta_{k,1} \qquad (25.13)$$

To solve Eq.(25.13) we define the generating function:

$$g(z) = \sum_{k=0}^{\infty} P(k) \times z^k \qquad (25.14)$$

Multiplying Eq.(25.13) by $z^k$ and summing over $k$ from 2 to infinity we get (taking into account Eq.(25.14)):

$$3g(z) = g^2(z) + z \times g(z) + z \qquad (25.15)$$

The solution of Eq.(25.15), which satisfies $P(0) = 0$, or $g(0) = 0$, is:

$$g(z) = \frac{3 - z - \sqrt{9 - 10 \times z + z^2}}{2} \qquad (25.16)$$

Eq. (25.16) can be rewritten as:

$$g(z) = \frac{3}{2} - \frac{z}{2} - \frac{3}{2}\sqrt{1-z} \times \sqrt{1 - \frac{z}{9}} \qquad (25.17)$$

The Taylor expansion of the square root is

$$\sqrt{1-X} = -\sum_{n=0}^{\infty} \alpha(n) X^n, \qquad (25.18)$$

where

$$\alpha(n) = \frac{\Gamma(n-1/2)}{2\sqrt{\pi}\Gamma(n+1)} \qquad (25.19)$$

After substituting Eq. (25.18) into Eq. (25.17) and using Eq.(25.4) we get:

$$P(k) = -\frac{3}{2} \sum_{n=0}^{k} \alpha(n) \times \alpha(k-n) \times \left(\frac{1}{9}\right)^n \qquad (25.20)$$

As the terms in Eq. (25.20) decrease exponentially with $n$ than for large $k$ one can replace the upper limit of summation with infinity and $\alpha(k-n)$ with $\alpha(k)$:

$$P(k) \propto -\frac{3}{2}\alpha(k)\sum_{n=0}^{\infty}\alpha(n)\times\left(\frac{1}{9}\right)^n =$$
$$\frac{3}{2}\sqrt{1-\frac{1}{9}} \times \alpha(k) = \frac{\Gamma(k-1/2)}{\sqrt{2\pi}\Gamma(k+1)}$$

Finally using the well known asymptotic of Gamma-function we get:

$$P(k) \propto \frac{1}{\sqrt{2\pi} \times k^{3/2}}$$

## 26. Distribution of genes

As we have seen, the branching process produces a power law in the distribution of the total number of descendants. There is another way to get a power law from the branching process. In 1964 Kimura and Crow [71] considered the following problem. There is a fixed size population and within it there are many alleles (alternative forms) of a particular gene. They assume that these alleles are selectively neutral, that is, the average offspring of individuals does not depend on which particular alleles of the gene they carry. With the probability $\alpha$ any individual gene can mutate creating a new allele. Kimura and Crow asked what the distribution of alleles in any given generation is.

Let us consider a model where the gene pool has constant size $N$. To form the next generation we $N$ times select a random gene from current generation pool and copy it to the next generation pool. With probability $\alpha$ each gene can mutate during the process of copying. An allele is equivalent to a family name in Bienaymé-Galton model. The average non-mutant offspring (which still carries family name) is equal to $\lambda = 1 - \alpha$. In the limit of large $N$ the offspring distribution is Poissonian. We denote as $N(m)$ the number of alleles represented $m$ times in the gene pool. The equilibrium distribution of $N(m)$ should satisfy the following self-consistency equation:

$$N(n) = \sum_{m=1}^{\infty} N(m) \frac{(\lambda m)^n}{n!} e^{-\lambda m} + \delta_{n,1} \alpha N \qquad (26.1)$$

In the limit of large $n$ the sum can be replaced with the integral:

$$N(n) = \frac{1}{n!} \int_0^{\infty} dm N(m) (\lambda m)^n e^{-\lambda m} \qquad (26.2)$$

In the case $\lambda = 1$ Eq.(26.2) has the solution $N(m) = C$, where $C$ is an arbitrary constant. Clearly, the integral becomes a gamma-function and the factorial in the denominator cancels out. However, this solution is, meaningless since the size of the gene pool, which is given by the equation

$$N = \sum_{m=1}^{\infty} m N(m) \qquad (26.3)$$

diverges.

In the case $\lambda < 1$, $N(m) = C$ is no longer a solution since the integral is equal to $C/\lambda$. However $N(m) = C/m$ is a solution. This solution is again meaningless because the size of the gene pool given by Eq.(26.3) again diverges. One can look for a solution of the form

$$N(m) = \frac{C}{m} \exp(-\beta m) \qquad (26.4)$$

After substituting Eq.(26.4) into Eq. (26.2) we get that $N(n)$ is given by Eq.(26.4) only with $\beta$ replaced with

$$\beta' = \ln(1 + \beta/\lambda) \qquad (26.5)$$

The self consistency equation for $\beta$ is thus

$$\beta = \ln(1 + \beta/\lambda). \qquad (26.6)$$

The obvious solution is $\beta = 0$ which gives us the previously rejected solution $N(m) = C/m$. It is also easy to see that this stationary solution is unstable. If $\beta$ slightly deviates from zero Eq.(26.5) gives us $\beta' = \beta/\lambda$. Since $\lambda < 1$ the deviation from the stationary shape will be bigger in the next generation. Another solution of Eq.(26.6) can be found by expansion of logarithm in Eq.(26.6) up to second order in $\beta$. It is $\beta \approx 2(1 - \lambda)$. One can show that it is stable. Thus we get

$$N(m) \approx \frac{C}{m} \exp(-2(1-\lambda)m) \qquad (26.7)$$

After substituting this into Eq.(26.3) we get

$$C \approx 2(1-\lambda)N \qquad (26.8)$$

Now we can estimate the number of distinct alleles in the gene pool:

$$D = \sum_{m=1}^{\infty} N(m) \approx 2\alpha N \ln\left(\frac{1}{\alpha}\right) \qquad (26.9)$$

The work of Kimura and Crow was based on the 1930 work of Wright [72] who considered a problem of a single type of mutation. So that in each generation there are $N$ genes, the fraction $q$ of which are mutant. With probability $u$ each mutant gene can mutate back to normal and each normal gene mutates into the only available mutant form with the probability $v$. Wright calculated the probability distribution of $q$. The problem is to that of Kimura and Crow since the distribution of a single mutant sampled over different generations should match the distribution of many mutants sampled in a single generation. The solution that we present is based on that given by Wright though it is a lot simpler. We at the very beginning considered the limit of large $N$ and got Poisson distribution and Gamma functions. Wright considered finite $N$ and consequently got Binomial distribution and Beta functions. If we take the limit of large $N$ (and, correspondingly, small $q$) and set $v$ to zero in the equation on the top of page 123 of Ref.[72], we recover Eq.(26.7). Note that we also need to divide Wright's $N$ by 2 since in his model $N$ is the number of individuals and each individual carries two sets of genes in two paired chromosomes.

In 1958 by Moran [43] proposed a variation of Wright's model. In his model not the whole generations of genes are updated at once, but genes die *one by one* at random and are replaced with new genes. We studied that model in Section 8.

Alternative solution [98] of the problem shows its connection to random walks in a potential. Suppose that the number of copies of an individual gene in the next generation has mean $\lambda$ and variance $\sigma^2$. If in current generation there is a large number $n$ of copies of given allele then the number of copies of this allele in the next generation comes from a normal distribution with mean $\lambda n$ and variance $\sigma^2 n$. The change in $n$ is

$$\Delta n = (\lambda - 1)n + \sqrt{n}\,\sigma \times z \qquad (26.10)$$

where $z$ is a normally distributed random number with zero mean and unit variance. The number of copies of given allele, $n$, performs a random walk, with the size of the step proportional to $\sqrt{n}$. Eq.(26.10) can be simplified by changing variable from $n$ to $x = \sqrt{n}$. Using Ito's formula [99], we get

$$\Delta x = \frac{(\lambda - 1)}{2} x - \frac{1}{8} \times \frac{1}{x} + \frac{\sigma}{2} z \qquad (26.11)$$

If not for the $\dfrac{1}{x}$ term we would get a well studied problem of Brownian motion in a harmonic potential [100]. Instead we get Brownian motion in the potential

$$U(x) = -\frac{(\lambda - 1)}{4} x^2 + \frac{1}{8} \times \ln(x). \qquad (26.12)$$

One can find probability density of $x$, by solving the corresponding Fokker-Planck equation. Alternatively it can be found as a Boltzmann distribution in the potential given by Eq. (26.12) at an appropriate temperature. The result is:

$$P(x) \sim \exp(-8U(x)) = \exp\left(\frac{2(\lambda - 1)x^2}{\sigma^2}\right)\frac{1}{x} \qquad (26.13)$$

As was pointed out in Ref.[7] "power laws are logarithmic Boltzmann laws."

The probability distribution of $n = x^2$ can be immediately found using Eq.(26.12):

$$P(n) \sim \exp\left(\frac{2(\lambda - 1)n}{\sigma^2}\right)\frac{1}{n} \qquad (26.13)$$

## 27. The Science of Self-Organized Criticality

In its mean-field version, Self Organized Criticality (SOC) [39] can be described as a branching process [41], [42]. Here the sand grains, which move during the original toppling, are equivalent to sons. This moved grains can cause further toppling, resulting in the motion of more grains, which are equivalent to grandsons, and so on. The total number of displaced grains is the size of the avalanche and is equivalent to total offspring in the case of a branching process. Size distribution of offspring is equivalent to distribution of avalanches in SOC. To be fair SOC is not completely reduced to a critical branching process: it has a built in dissipative mechanism (sand grains fall from the edges), which tunes the branching process into a critical state.

Let us consider a mean field version of SOC on a lattice with coordination number $k$. Then we have the following generation function for the offspring probability:

$$f(z) = 1 - p(k) + p(k)z^k \qquad (27.1)$$

We should substitute Eq.(27.1) into Eq.(19.2) to obtain the distribution of the sizes of the avalanches. In the case $k = 2$ we can easily solve the resulting quadratic equation. In the general case we should use Eq.(A2) to get:

$$P(n) = \frac{1}{2\pi \times i} \times$$

$$\oint dg \, \frac{\left(1 - p(k) + p(k) \times g^k\right)^n}{n g^n} =$$

$$= \frac{(1-p(k))^{mk+1-m} (p(k))^m}{mk+1} \binom{mk+1}{m} =$$

$$\frac{(1-p(k))\left(p(k)(1-p(k))^{k-1}\right)^m}{mk+1} \frac{(mk)!}{m!(m(k-1)+1)!}$$

where $n = mk + 1$.

For large $n$ we obtain the following asymptotic for the distribution of avalanche sizes:

$$P(n) =$$

$$\frac{(1-p(k))k^2}{\sqrt{2\pi}(k-1)^{3/2}} \frac{1}{n^{3/2}} \left( k \times p(k) \left( \frac{1-p(k)}{1-1/k} \right)^{k-1} \right)^{n/k} \quad (27.2)$$

In the critical regime, when $p(k) = 1/k$, Eq. (27.2) gives

$$P(n) = \frac{k}{\sqrt{2\pi}(k-1)^{1/2}} \frac{1}{n^{3/2}}$$

## 28. Yule's model as a continuous-time Branching Process

Finally, we would like to show that Yule's model can be viewed as a special kind of branching process. This connection was pointed out by Harris (see p.105 of Ref. [38]).

We want to compute the probability ($p_n(t)$) that a genus of age $t$ has exactly $n$ species in it. Recall that the density function for the age of a randomly picked genus is given by the exponential distribution: $f(t) = g \exp(-gt)$. As shown before, using the age distribution of genus and the distribution of the total number of species in a genus of a given age, one obtains the distribution for the size of a randomly picked genus; see Equation (2.5).

The assumption that each species independently mutates at the rate of $s$ is equivalent to saying that the mutation process of a single species is a Poisson process. Thus, if we consider the first species in a genus, then the number of its "children" (i.e., the number of species that results from mutations of this original species only) in an interval t, is a Poisson random variable with mean $st$. Now suppose one of these children was born at time $t_1$, then the number of its children (hence, grand-children of the original species) is a Poisson random variable with mean $s(t - t_1)$. Hence, the evolution of a genus can be modeled as a *continuous-time branching process*, and as shown in the following, one can use the generating function approach used in the context of discrete-time branching processes to calculate the distributions of the number of species in a particular generation, *as well as*, the total number of species in a genus of a given age.

Note that for a standard branching process, the generating function for the distribution of the number of grandkids is easily computed by realizing that generating function is an expected value:

$$f_2(z) = E[z^{X_2}] = \sum_{k=0}^{\infty} P(X_1 = k) E[z^{X_2} \mid X_1 = k].$$

Now, given $X_1 = k$, $X_2$ is the sum of $k$ iid Poisson random variables, and recalling that the generating function of the sum of two independent random variables is the product of their respective generating functions, we get $E[z^{X_2} \mid X_1 = k] = (f(z))^k$, where

$$f(z) = \sum_{k=0}^{\infty} P[X_1 = k] z^k$$ is the generating function

for the number of kids of the original starting node. Thus,

$$f_2(z) = \sum_{k=0}^{\infty} P(X_1 = k) E[z^{X_2} \mid X_1 = k] =$$

$$\sum_{k=0}^{\infty} P(X_1 = k)(f(z))^k = f(f(z)).$$

We will use the preceding arguments, based on conditional expectations, repeatedly in the following, where the age of a species will come in as an extra variable.

In the case of continuous-time branching process, a node keeps on producing more offspring with time. For a fixed time, the number of children is still a Poisson random variable. The distribution of children of the original species in a genus of age $t$, $X_1(t)$, is

$P[X_1(t) = k] = e^{-st}(st)^k / k!$,

and the corresponding generating function is

$$f(z,t) = E[z^{X_1(t)}] = \sum_{k=0}^{\infty} P(X_1(t) = k)z^k = \exp(st(z-1)).$$

Now, since in a Poisson process the intervals between the births of two successive children are *iid* exponential random variables, the computation of the conditional distribution of the birth times of the children, given that exactly k children were born in a time interval t, turns out to be surprisingly simple, and the related result can be stated as follows:

*Given that k events have occurred in the interval $(0,t)$, the times $S_1, S_2, \ldots, S_k$ at which events occur, considered as unordered random variables, are distributed independently, and uniformly, in the interval $(0,t)$.*

Thus, given that the original species has k children, the distribution of its grandchildren will be the same as when the birth times of its k children are picked independently and uniformly in the interval $(0,t)$. If $f_2(z,t)$ is the generating function of the distribution of the grandkids, then one can easily verify following the same arguments as in the standard model of branching process that

$$f_2(z,t) = \sum_{k=0}^{\infty} P(X_1(t) = k)\left(\frac{1}{t}\int_0^t f(z,(t-y))dy\right)^k,$$

where as computed before, $f(z,t) = \exp(st(z-1))$. Hence, we get

$$f_2(z,t) = \sum_{k=0}^{\infty} P(X_1(t) = k)\left(\frac{e^{st(z-1)} - 1}{st(z-1)}\right)^k =$$

$$f\left(\frac{e^{st(z-1)} - 1}{st(z-1)}, t\right) = \exp\left(\frac{e^{st(z-1)} - (1 - st(z-1))}{(z-1)}\right).$$

In general, if $f_k(z,t)$ is the generating function of the distribution of the number of species in

generation k, then following the same argument as in the preceding, we get

$$f_k(z,t) = \sum_{k=0}^{\infty} P(X_1(t) = k)\left(\frac{1}{t}\int_0^t f_{(k-1)}(z,(t-y))dy\right)^k =$$

$$f\left(\frac{1}{t}\int_0^t f_{(k-1)}(z,(t-y))dy, t\right),$$

where $f_1(z,t) = f(z,t) = \exp(st(z-1))$. This is a generalization of the standard branching process recursion: $f_k(z) = f(f_{(k-1)}(z))$, where $f_1(z) = f(z)$.

In a standard branching process, the generating function, $g(z)$, corresponding to the distribution of the total number of children over all generations, is derived from the straightforward self-consistency equation $g(z) = zf(g(z))$. The same arguments, augmented with the reasoning in the preceding discussion, can be used to show that for our continuous-time branching process the self-consistency equation is

$$g(z,t) = z\left(\sum_{k=0}^{\infty} P(X_1(t) = k)\left(\frac{1}{t}\int_0^t g(z,t-y)dy\right)^k\right) =$$

$$zf\left(\frac{1}{t}\int_0^t g(z,y)dy, t\right) = z\exp\left(st\left(\left(\frac{1}{t}\int_0^t g(z,y)dy\right) - 1\right)\right).$$

So we want to solve for $g(z,t)$ that satisfies the following integral equation:

$$g(z,t) = z\exp\left(st\left(\left(\frac{1}{t}\int_0^t g(z,y)dy\right) - 1\right)\right).$$

Differentiating both sides with respect to time, and noting that $\frac{\partial}{\partial t}\int_0^t g(z,y)dy = g(z,t)$, we get

$$\frac{\partial}{\partial t}g(z,t) = sg(z,t)(g(z,t) - 1),$$

where $g(z,0) = z$ (i.e., at $t = 0$, there is only the parent node in the chain). The above equation, with the given initial condition is easy to solve and we get

$$\ln\left(1 - \frac{1}{g(z,t)}\right) = st + \ln\left(1 - \frac{1}{z}\right).$$

Simplifying, we get

$$g(z,t) - 1 = g(z,t)\left(1 - \frac{1}{z}\right)\exp(st),$$

which yields

$$g(z,t) = \frac{\exp(-st)z}{1 - (1 - \exp(-st))z} =$$

$$\sum_{k=1}^{\infty} \exp(-st)(1 - \exp(-st))^{k-1} z^k.$$

Hence, by definition,

$$p_n(t) = \exp(-st)(1 - \exp(-st))^{(n-1)},$$

for all $n \geq 1$.

Note that in Yule's model we actually have *two nested continuous-time Branching processes*: The first one corresponds to the evolution of species within a given genus, and we computed the size distribution of a genus with a given age. The second one corresponds to the evolution of genus, i.e., each genus gives rise to new ones at the rate of $g$. One could apply the same framework as before, and thus, the distribution of the number of genus at time $T$ is given as:

$$P[no.\,of\,genus = l \quad at \quad time \quad T] = q_n(T) =$$

$$\exp(-gT)(1 - \exp(-gT))^{(l-1)}.$$

## 29. Re-inventing Willis

We summarized the re-inventions, described in this paper, in Table 29.1. We treat Simon's and Yule's models as different things, because they use different mathematical approaches (alternative ways to America). We count it as a re-discovery when the same America is discovered in the same way. Even with this restriction, almost everything appears to be re-discovered twice.

**Table 29.1.** Summary of all re-inventions.

| Phenomenon | Discovered | Re-discovered | | | | | |
|---|---|---|---|---|---|---|---|
| Branching process | Bienaymé, Cournot (1845-1847) | Galton & Watson (1873-1875) | Fisher (1922) | Erlang, Steffensen, Christensen (1929-1930) | Shockley & Pierce (1938) Woodward (1947) | Flory (1941) | Hawkins & Ulam (1944) |
| Yule's process | Yule (1925) | Fermi (1949) | Huberman and Adamic (1999) | | | | |
| Simon's process | Simon (1955) | Günter et al (1992) | Barabasi & Albert (1999) | | | | |
| Power law of word frequencies | Estoup (before 1916) | Condon (1928) | Zipf (1935) | | | | |
| Power law of scientific citing | Price (1965) | Silagadze (1997) | Redner (1998) | | | | |
| Champernowne's process | Champernowne (1953) | Levy and Solomon (1996) | | | | | |
| Urn model | Markov (1907) | Eggenberger and Polya (1923) | | | | | |
| Random graph | Flory (1941) | Erdos & Renyi (1960) | | | | | |

Such pattern of re-discoveries is not limited to the particular scientific island, which was the focus of our attention. The statistician Stephen M. Stigler even formulated the law: "No scientific discovery is named after its original discoverer." See the chapter "Stigler's law of eponymy" in Ref. [82]. The sociologist Stanislav Andreski in his book "Social sciences as sorcery" [83] wrote, "Rediscovering America is one of the most popular occupations among practitioners of the social sciences." We conclude that the scientists are busy with re-inventing Willis most of the time.

## Appendix

Suppose that $z = F(g)$, where $F(g)$ is an analytic function, such that $F(0) = 0$ and $F'(0) \neq 0$. We want to find the expansion of some analytic function $\Phi(g)$ in powers of $z$:

$$\Phi(g(z)) = \Phi(0) + \sum_{n=1}^{\infty} \Phi_n z^n. \tag{A1}$$

We can use Cauchy formula to find $\Phi_n$:

$$\Phi_n = \frac{1}{2\pi \times i} \oint dz \frac{\Phi(g)}{z^{n+1}} =$$

$$\frac{1}{2\pi \times i} \oint dg \left(\frac{dz}{dg}\right) \frac{\Phi(g)}{z^{n+1}} =$$

$$\frac{1}{2\pi \times i} \oint dg \frac{F'(g)\Phi(g)}{(F(g))^{n+1}} =$$

$$\frac{1}{2\pi \times i} \oint dg \left(\frac{\Phi'(g)}{n(F(g))^n} - \frac{d}{dg}\left(\frac{\Phi(g)}{n(F(g))^n}\right)\right)$$

When $F(g)$ and $\Phi(g)$ are single-valued functions the second term gives zero after integration and we have

$$\Phi_n = \frac{1}{2\pi \times i} \oint dg \frac{\Phi'(g)}{n(F(g))^n} \tag{A2}$$

When $F'(g) \neq 0$ the integrand has a pole of a degree up to $n$ and we can rewrite

$$\frac{\Phi'(g)}{(F(g))^n} = \frac{\Phi'(g)(g/F(g))^n}{g^n}$$

where the numerator is an analytic function. The non-vanishing after contour integration part is given by the $(n\text{-}1)$th term in its Taylor expansion, which is

$$\frac{1}{(n-1)!} \frac{d^{n-1}}{dg^{n-1}} \left(\Phi'(g)\left(\frac{g}{F(g)}\right)^n\right)\Bigg|_{g=0}$$

After substituting this into Eq. A2 we get:

$$\Phi_n = \frac{1}{n!} \frac{d^{n-1}}{dg^{n-1}} \left(\Phi'(g)\left(\frac{g}{F(g)}\right)^n\right)\Bigg|_{g=0} \tag{A3}$$

which is known (together with Eq. A1) as Lagrange expansion. In some of the applications we are dealing with in the paper it is more convenient to use Eq A2 than Eq A3.